\documentclass[12pt,a4paper]{article}
\usepackage[truedimen,margin=30mm]{geometry} 

\usepackage{mathrsfs}
\usepackage{amssymb}
\usepackage{amsmath}
\usepackage{ascmac}
\usepackage{amsthm}
\usepackage{bm}
\usepackage[dvipdfmx]{graphicx}
\usepackage{natbib}
\usepackage{setspace}
\usepackage{comment}

\usepackage{color}
\usepackage{url}
\usepackage{times}

\usepackage{titlesec}
\titleformat*{\section}{\large\bfseries}
\titleformat*{\subsection}{\it}
\titleformat*{\subsubsection}{\it}

\newtheorem{prp}{Proposition}

\newcommand{\bbeta}{\boldsymbol{\beta}}

\newcommand{\bx}{\boldsymbol{x}}

\title{{\bf The Group R2D2 Shrinkage Prior for Sparse Linear Models with Grouped Covariates}\footnote{\today}}

\author{}
\date{}

\begin{document} 

\maketitle
\doublespacing

\vspace{-1.7cm}
\begin{center}
{\large 
Eric Yanchenko\footnote{Human \& AI Center, Akita International University, Akita, Japan. \url{eyanchenko@aiu.ac.jp}}, 
Kaoru Irie\footnote{Faculty of Economics, The University of Tokyo, Tokyo, Japan} and 
Shonosuke Sugasawa\footnote{Faculty of Economics, Keio University, Tokyo, Japan}
}
\end{center}

\vspace{0.5cm}
\begin{center}
{\bf \large Abstract}
\end{center}

\noindent
Shrinkage priors are a popular Bayesian paradigm to handle sparsity in high-dimensional regression. Still limited, however, is a flexible class of shrinkage priors to handle grouped sparsity, where covariates exhibit some natural grouping structure. This paper proposes a novel extension of the $R^2$-induced Dirichlet Decomposition (R2D2) prior to accommodate grouped variable selection in linear regression models. The proposed method, called the gR2D2 prior, employs a Dirichlet prior distribution on the coefficient of determination for each group, allowing for a flexible and adaptive shrinkage that operates at both group and individual variable levels. We present several theoretical properties of this proposed prior distribution while also developing a Markov Chain Monte Carlo algorithm. Through simulation studies and real-data analysis, we demonstrate that our method outperforms traditional shrinkage priors in terms of both estimation accuracy, inference and prediction.

\bigskip\noindent
{\bf Key words}: Dirichlet distribution; grouped variable selection; logistic normal distribution

\noindent
{\bf Acknowledgments}: This work is supported by the Japan Society for the Promotion of Science KAKENHI (grant numbers: 20H00080, 21H00699, 24K00244, 24K00175, 24K21420 and 24K22613). The authors are also grateful for reviewer and editor comments which greatly improved the quality of the manuscript.

\newpage
\section{Introduction}

Shrinkage priors are a popular Bayesian tool to handle sparsity in high-dimensional regression settings, and ever since the Horseshoe prior \citep{carvalho2009handling}, they have garnered significant research attention. One area, however, where they have yet to be fully exploited is in grouped regression. In this setting, the covariates possess some natural grouping structure which can potentially be leveraged to improve inference and prediction. 

To fix ideas, we give three examples where grouping structure naturally arises. First, consider a neuroimaging application where the response variable $\boldsymbol{y}$ is the change in brain activity for each neuron after some stimulus \citep[e.g.,][]{denti2023horseshoe}. If the number of neurons is large, we may endow the mean vector $\bbeta$ with a shrinkage prior to encode the {\it a priori} belief that many neurons are unaffected by the stimulus, while some may show a large change. Since different regions of the brain, based on either brain function or spatial location, provide a natural grouping structure, leveraging this information in the prior distribution of the mean vector can improve the inference on $\bbeta$ as compared to a vanilla shrinkage prior.

Next, network-valued data also provide opportunities for group shrinkage priors. For example, influence maximization (IM) \citep{kempe2003maximizing, li2018influence} is the problem of activating a small set of seed nodes to maximize the influence spread on the network. Recently, \cite{yanchenko2023bopim} proposed a Bayesian optimization solution to the IM problem which can leverage shrinkage priors, where each node corresponds to a coefficient $\beta_j$. Since networks often exhibit community structure with nodes clustered together \citep{girvan2002community, fortunato2010community}, group shrinkage priors can improve performance.

Finally, in economics, covariate grouping may arise from many dummy variables that correspond to a single covariate. For example, a firm's age may be a useful predictor of their payment behavior \citep{garavaglia1998smart}. Although age is a continuous variable, it can be advantageous to transform it into discrete levels, i.e., 0-2 years old, 3-7 years old, etc.  While different coefficients now represent each age bracket, they all correspond to the same original variable (firm's age), leading to a natural grouping. In addition to grouping structure that arises from domain knowledge of the predictor variables, there may also be a known structure that can be leveraged in, e.g., generalized additive models \citep{xu2016bayesian}. Here, the basis functions corresponding to each spline can be treated as a group. By using shrinkage priors with grouping, if an entire group of covariates are unimportant, their effect estimates will be simultaneously shrunk.

As there are many relevant applications for grouped-shrinkage regression, there have been several contributions to the literature. First, many articles are based on the Bayesian group LASSO \citep{raman2009bayesian, casella2010penalized, xu2015bayesian}. As \cite{tibshirani1996regression} notes in his seminal paper, the maximum likelihood estimate of the LASSO can be thought of as the Bayesian posterior mode of the regression coefficients with Laplace prior distribution. Thus, endowing the $\beta_j$'s with a Laplace prior was coined the Bayesian LASSO. This can easily be extended to the grouped regression setting. If we let $\bbeta=(\bbeta_1^{\top},\dots,\bbeta_G^{\top})^{\top}$ where $\bbeta_g$ are the coefficients for group $g\in\{1,\dots,G\}$, then a prior of the form
$$
\pi(\bbeta_g)
\propto \exp\left\{-\frac{\lambda}{\sigma}\|\bbeta_g\|_1\right\},
$$
where $||\cdot||_1$ is the $L_1$-norm, is the multivariate extension of the Laplace distribution and induces a grouped LASSO prior on $\bbeta$ \citep{xu2015bayesian}. Additionally, \cite{xu2016bayesian} proposed an extension of the Horseshoe prior to the grouped setting. 

More recently, \cite{boss2023group} proposed the group inverse-gamma gamma (GIGG) prior as a novel global-local shrinkage prior for the grouped regression setting. In this method, the variance of each regression coefficient is comprised of three components. The first component is a global variance, endowed with a Half-Cauchy prior distribution, which is shared by all coefficients and controls the overall variability. The second is the group-specific variance parameter to shrink the relative contribution of each group. The $G$ group-specific variances are given gamma prior distributions. The last is a local term, with inverse-gamma prior distribution, to allow for individual coefficients to exhibit a strong signal.

While hitherto not applied to the grouped regression setting, the R2D2 prior \citep{zhang2022bayesian} has shown great promise in sparse regression settings thanks to its excellent empirical and theoretical properties. Setting a prior distribution on a Bayesian coefficient-of-determination ($R^2$) induces distributions on the global and local variance components. In particular, if $\beta_j\sim\mathsf{Normal}(0,\sigma^2\phi_jW)$ and $\sum_{j=1}^p \phi_j=1$, then the Bayesian $R^2$ is 
$$
R^2=\frac{W}{W+1}.
$$
where $W$ is the global variance parameter. Thus, a beta prior distribution on $R^2$ induces a Beta-prime distribution on $W$, the global variance parameter, while the local parameters $\phi_j$'s apportion the variance to each coefficient. Originally proposed in the high-dimensional linear regression setting, the approach has been extended to generalized linear models \citep{aguilar2023intuitive, yanchenko2024r2d2}, spatial models \citep{yanchenko2024spatial}, auto-regressive models \citep{kohns2025arr2}, ordinal models \citep{yanchenko2025pseudo} and more \citep{gan2022graphical}.

In this work, we leverage the R2D2 paradigm to develop a novel prior framework for the high-dimensional, grouped linear regression setting. We begin by deriving both a global measure of $R^2$ to quantify the overall model fit, as well as group-specific $R^2$'s to measure the relative contribution of each group of coefficients to the overall fit. These measures extend the global-local shrinkage framework to include group-level sources of variation. Placing a Dirichlet distribution on the group-specific and global $R^2$ induces (dependent) Beta-prime distributions on the global and group-specific variance parameters. Both the Dirichlet distribution and logistic normal distribution \citep{aguilar2024generalized} are considered to allocate the variance to the local components. Finally, we derive some theoretical properties of this prior distribution, as well as an efficient MCMC sampler for posterior approximation. The proposed method is shown to outperform the current state-of-the-art on simulated and real-world datasets. 

The layout of the paper is as follows. We start by presenting the proposed method in Section \ref{sec:method} and discuss its computational properties in Section \ref{sec:comp}. We demonstrate its performance on simulated and real-world data in Sections \ref{sec:sim} and \ref{sec:real}, respectively, while Section \ref{sec:conc} shares concluding thoughts.

\section{Group R2D2 Prior}\label{sec:method}
\subsection{Linear regression model}
We consider the following linear regression model: 
\begin{equation}\label{eq:model}
Y_i=\bx^{\top}_i\bbeta+\varepsilon_i, \ \ \ \ \ \varepsilon_i\stackrel{\text{iid}}{\sim} \mathsf{Normal}(0, \sigma^2), \ \ \ \ \ i=1,\ldots,n,
\end{equation}
where $Y_i$ is the response, $\bx_i$ is a $p$-dimensional vector of explanatory variables, $\bbeta$ is the $p$-dimensional coefficient vector and $\epsilon_i$ is the error term.
For simplicity, we assume that the responses are centered, so that the model in (\ref{eq:model}) does not require an intercept term.

We consider the scenario where there is a known grouping structure in the regression coefficients. The groups could correspond to, e.g., different brain regions, communities in networks, or basis functions. Specifically, let $\bx_g$ and $\bbeta_g=(\beta_{g1},\dots,\beta_{gp_g})^{\top}$ be the explanatory variables and regression coefficients, respectively, corresponding to group $g\in\{1,\dots,G\}$, where $G$ is the total number of groups, $p_g$ is the number of variables in group $g$, and $\sum_g p_g=p$. Then, $\bx$ and $\bbeta$ can be expressed as $\bx=(\bx_1^{\top},\dots,\bx_G^{\top})^{\top}$ and $\bbeta=(\bbeta_1^{\top},\dots,\bbeta_G^{\top})^{\top}$.
For regression coefficients $\beta_{gj}$, we assign a prior having zero mean and finite variance $\sigma^2\lambda_{gj}$, or $\beta_{gj}|\lambda_{gj}, \sigma^2 \sim (0,\sigma^2\lambda_{gj})$ independently for $j=1,\dots,p_g$ and $g=1,\dots,G$. Specifically, we use the double-exponential distribution as the prior for $\beta_{gj}$ in what follows. The goal is to find the distribution of these parameters which induces a prior distribution with desirable shrinkage properties.

\subsection{Grouped R2 and priors on group-level variance parameters}
We derive the grouped regression shrinkage prior following the argument of the R2D2 prior \citep{zhang2022bayesian}. 
Let $W_g=\sum_{j=1}^{p_g}\lambda_{gj}$ for $g=1,\ldots,G$, and $W=\sum_{g=1}^G W_g=\sum_{j=1}^{p_g}\lambda_{gj}$, i.e., $W_g$ quantifies the variation in group $g$ while $W$ quantifies the overall variation in the model.
Moreover, without loss of generality, we assume that $\mbox{E}(\bx)={\bf 0}_p$ and $\mbox{Var}(\bx)=\Sigma_x$ where $\mathbf{0}_n$ is the $n$-dimensional zero vector and $\Sigma_x$ has one's on the diagonal. We further assume that $\bx$ and $\bbeta$ are independent.
Then, for a random variable $Y=\bx^\top \bbeta +\varepsilon$, it follows that 
$$
\mathrm{Var}(Y)=\mathrm{Var}(\bx^{\top}\bbeta) + \sigma^2
=\sigma^2\sum_{g=1}^G\sum_{j=1}^{p_g}\lambda_{gj}+\sigma^2
$$
Hence, we define the global coefficient of determination as 
\begin{equation*}
R^2 = \frac{\mathrm{Var}(\bx^{\top}\bbeta)}{\mathrm{Var}(Y)} = \frac{ \sum _{g=1}^G \sum _{j=1}^{p_g} \lambda _{gj} }{ \sum _{g=1}^G \sum _{j=1}^{p_g} \lambda _{gj} + 1 } = \frac{ W }{ W +1},
\end{equation*}
and its group-specific version as  
\begin{equation*}
R^2_g = \frac{\mathrm{Var}(\bx_g^{\top}\bbeta_g)}{\mathrm{Var}(Y)} = \frac{ \sum _{j=1}^{p_g} \lambda _{gj} }{ \sum _{g'=1}^G \sum _{j=1}^{p_{g'}} \lambda _{g'j} + 1 } = \frac{ W_g}{ W +1}, \ \ \ \ g=1,\ldots,G.
\end{equation*}
Please see the Supplemental Materials for further details on these derivations. This definition ensures $\sum_{g=1}^G R_g^2=R^2$. Therefore, $R^2$ is interpreted as the proportion of total variation in the response explained by all coefficients $\bbeta$, while $R^2_g$ is the proportion of total variation in the response explained by the coefficients only in group $g$, $\bbeta_g$. 
In this way, the larger $R^2_g$ is relative to $R^2$, the more the coefficients in group $g$ account for the variation explained by the model. Thus, $R^2_g$ is a relative measure of group $g$'s importance.

Given the definition of $R^2$ and $R^2_g$, we must now set prior distribution for these parameters. In \cite{zhang2022bayesian}, the authors set $R^2\sim\mathsf{Beta}(a,b)$. In the grouped setting, however, we notice that the group-specific $R^2_g$ is affected by $R^2_{g'}$ for $g'\neq g$. 
Thus, we propose a natural extension to the Dirichlet distribution: 
\begin{equation}\label{eq:dist}
(R^2_1,\dots,R_G^2,1-R^2)\sim\mathsf{Dirichlet}(a_1,\dots,a_G,b).
\end{equation}
This marginally induces $R^2\sim\mathsf{Beta}(\sum_{g=1}^G a_g, b)$ and $R^2_g\sim\mathsf{Beta}(a_g,b+\sum_{g'\neq g} a_{g'})$. 
The following proposition states sufficient conditions such that this joint prior on $R^2_g$ and $R^2$ is realized, with proof in the Supplemental Materials. Throughout the remainder of this work, we use the shape-scale parameterization of the gamma distribution, i.e., if $X\sim\mathsf{Gamma}(\alpha,\theta)$, then $\mathrm{E}(X)=\alpha\theta$.
\begin{prp} \label{prp:r2d2}
If $W_g|\tau^2\sim\mathsf{Gamma}(a_g,\tau^2)$ and $\tau^2\sim\mathsf{InvGamma}(b,1)$, all independently, then $(R^2_1,\dots,R^2_G,1-R^2)\sim\mathsf{Dirichlet}(a_1,\dots,a_G,b)$.
\end{prp}

Proposition 1 clarifies which part of the model must be specified so that the $R^2_g$'s are Dirichlet distributed. This result also implies a Beta-prime distribution on the global variance, i.e., $W|\tau^2\sim\mathsf{Gamma}(\sum_{g=1}^G a_g,\tau^{2})$, or equivalently, $W\sim\mathsf{BetaPrime}(\sum_{g=1}^G a_g, b)$.  Moreover, the group-specific variances $W_g$'s are marginally Beta-prime distributed, but mutually dependent via $\tau^2$.

To identify which part of the model is unspecified by Proposition~\ref{prp:r2d2}, we follow the notation of \cite{zhang2022bayesian} and introduce weight parameter $\phi _{gj}$ so that $\lambda_{gj}= \phi_{gj} W_g$, where $\boldsymbol{\phi}_g= (\phi _{g1},\dots ,\phi _{gp_g})$ lies on the $(p_g{-}1)$-dimensional simplex. Then it holds that $W_g = \sum _{j=1}^{p_g}\lambda_{gj}$ for $g=1,\dots,G$. While Proposition~\ref{prp:r2d2} yields the prior distribution for $W_g$, the prior for $\boldsymbol\phi_g$ is arbitrary. The choice of priors for $\boldsymbol\phi_g$ is discussed in Section~\ref{subsec:phi}.

\subsection{The group R2D2 prior}
To complete the prior for $\bbeta$, the class of distributions for $\beta_{gj}|\lambda_{gj}, \sigma^2$ must be specified. 
While any distribution with support on the entire real line and finite variance is possible, we adopt the double exponential (Laplace) distribution, as used in the original R2D2 prior, to enhance the shrinkage effects in the posterior. 
The resulting prior of $\bbeta$ is given by 
\begin{equation*}
\beta_{gj}|\phi_{gj}, W_g,\sigma^2\sim \mathsf{DE}\left(\sigma^2 \phi_{gj} W_g  \right), \quad W_g|\tau^2\sim\mathsf{Gamma}(a_g,\tau^{2}), \quad \tau^2\sim\mathsf{InvGamma}(b,1),
\end{equation*}
where $\mathsf{DE}(\lambda)$ denotes the double exponential distribution with variance $\lambda$. We also let $\sigma^2\sim\mathsf{InvGamma}(n_0,d_0)$.\footnote{In all experiments, we set $n_0=d_0=1$.}
Using the scale mixture representation of the double exponential distribution \citep{andrews1974scale,west1987scale,park2008bayesian}, $\beta _{gj}$ becomes conditionally Gaussian with exponentially-distributed latent variable $\psi _{gj}$ as:
\begin{equation}\label{eq:R2D2}
\begin{split}
&\beta_{gj}|\psi_{gj},\phi_{gj}, W_g, \sigma^2 \stackrel{\text{ind.}}{\sim} \mathsf{Normal}\left(0, \frac12\sigma^2\psi_{gj} \phi_{gj}W_g\right)\\
&W_g|\tau^2 \stackrel{\text{ind.}}{\sim} \mathsf{Gamma}(a_g,\tau^2), \ \ \ \ \ 
\tau^2\sim\mathsf{InvGamma}(b,1), \ \ \ \ \psi_{gj}\stackrel{\text{ind.}}{\sim}\mathsf{Exp}\Big(\frac12\Big), 
\end{split}
\end{equation}
We call the class of priors for $\beta _{gj}$ the {\it group R2D2 (gR2D2) prior}. To reiterate, Proposition~\ref{prp:r2d2} yields the prior distributions for $W_g$'s and $\tau^2$ such that the grouped $R^2$'s are Dirichlet distributed, but the prior for weights $\phi _{gj}$ are arbitrary.

\subsection{Priors on local variance parameters} \label{subsec:phi}
To complete the prior specification, we must set the distribution on the local variance-allocation parameters, $\boldsymbol\phi_g$. We consider two choices: {\it Dirichlet distributions} and {\it logistic-normal distributions}. We denote these group R2D2 priors as {\it gR2D2-D} and {\it gR2D2-L}, respectively.  

\subsubsection*{Dirichlet priors (gR2D2-D)}

In previous works \citep{zhang2022bayesian, yanchenko2024r2d2}, a Dirichlet prior is used for the weight vector, i.e., $\boldsymbol{\phi}_g \sim \mathsf{Dirichlet} (a_{g1},\dots ,a_{gp_g})$.  Typically, the Dirichlet shape parameters are constrained so that $a_g = \sum _{j=1}^{p_g} a_{gj}$ \citep[e.g.][]{zhang2022bayesian}. This particular choice has been preferred for computational reasons, as samples can easily be drawn from the full conditional distribution. It also results in the within-group (conditional) independence, however, in the sense that
\begin{equation*}
\phi_{gj} W_g | \tau^2 \stackrel{\text{ind.}}{\sim}  \mathsf{Gamma}(a_{gj},\tau^2), \qquad j=1,\dots,p_g,
\end{equation*}
due to the gamma-Dirichlet relationship (see Proposition 5 of \cite{zhang2022bayesian}). In this study, we do not enforce this constraint on the hyper-parameters $a_g$. The Dirichlet distribution induces negative correlation between the $\phi_{gj}$'s, as pointed out in \cite{aguilar2024generalized}, which motivates considering other distributions.

\subsubsection*{Logistic-normal priors (gR2D2-L)}

We also consider the prior proposed in \cite{aguilar2024generalized} where the authors model $\phi_{gj}$ through a logistic-normal distribution. Specifically, for real-valued latent variable $\eta _{gj}$, we define
$$
\phi_{gj}=\frac{\exp(\eta_{gj})}{1+\sum_{j'=1}^{p_g-1}\exp(\eta_{gj'})}, \ \ \ \ \ j=1,\ldots,p_g-1,
$$
where $(\eta_{g1},\ldots,\eta_{g,p_g-1})\sim \mathsf{Normal}(\mathbf{0}_{p_g-1}, \Sigma_g)$ and $\Sigma_g$ is a fixed variance-covariance matrix. 
The joint distribution of $\boldsymbol\phi_g$ is then given by 
$$
p(\boldsymbol\phi_g)= |2 \pi \Sigma_g|^{-1/2}\left(\prod_{k=1}^{p_g} \phi_{gk}\right)^{-1} \exp \left\{-\frac{1}{2}{\rm LR}(\phi_g)^{\top} \Sigma_g^{-1}{\rm LR}(\phi_g)\right\},
$$
where $\mathrm{LR}(\cdot)$ is the log-ratio transformation:
$$
{\rm LR}(\boldsymbol\phi_g)
\equiv
\left(\log \left(\frac{\phi_{g1}}{\phi_{g,p_g}}\right), \ldots, \log \left(\frac{\phi_{g,p_g-1}}{\phi_{g,p_g}}\right)\right) \in \mathbb{R}^{p_g-1}.
$$
We say that $\boldsymbol\phi_g$ has a logistic normal distribution and write $\boldsymbol\phi_g\sim\mathsf{LogisticNormal}(\mathbf{0}_{p_g-1},\Sigma_g)$. 
This class of distributions could provide more flexibility in modeling  $\boldsymbol\phi_g$, but requires a Metropolis-Hastings step to generate posterior samples.

\subsection{Hyperparameters}\label{sec:hyper}
In this sub-section, we discuss several choices for setting hyperparameter values. The proposed gR2D2 priors include hyperparameters $a_g$ and $b$ which are common in gR2D2-D and gR2D2-L, and control the shape of Dirichlet prior for the group-wise coefficient of determination, as discussed in Proposition~1. Additionally, we must set $a_{gj}$ (only in gR2D2-D) and $\Sigma_g$ (only in gR2D2-L).


\subsubsection*{Empirical Bayes approach} 
We first discuss an Empirical Bayes approach to setting the values of $a_g$.
For a group-specific weight $w_g$, satisfying $\sum_{g=1}^G w_g=1$, we set $a_g = a w_g$, which results in the prior distribution for the global $R^2$ as $R^2\sim\mathsf{Beta}(a, b)$, as in the original R2D2 prior.
For specification of the weight $w_g$, we can employ a data-dependent approach. 
Under this parametrization, $\mathrm{E}[R_g^2]=aw_g/(a+b)=w_g \mathrm{E}[R^2]$, so $w_g$ is regarded as a ratio of the group-specific $R_g^2$ to the global $R^2$. 
Then, 
$$
w_g=\frac{\mathrm{E}[R_g^2]}{\mathrm{E}[R^2]}
=\frac{\mathrm{Var}(\bm{x}_g^{\top}\boldsymbol\beta_g)}
{\sum_{g'=1}^G \mathrm{Var}(\bm{x}_{g'}^{\top}\boldsymbol\beta_{g'})}.
$$
To obtain the point estimates of $w_g$, we first fit a linear regression model with only the $g$th covariates, $\bm{x}_g$, to obtain point estimates $\hat{\boldsymbol\beta}_g$. This yields a point estimate of $w_g$ as 
$$
\hat{w}_g=\frac{\hat{\boldsymbol\beta}_g^\top  \boldsymbol{V}_g\hat{\boldsymbol\beta}_g}
{\sum_{g'=1}^G\hat{\boldsymbol\beta}_{g'}^\top  \boldsymbol{V}_{g'}\hat{\boldsymbol\beta}_{g'}},
$$
where $\boldsymbol{V}_g$ is the $p_g \times p_g$ sample covariance matrix of $\boldsymbol{x}_g$.
As a default choice of $(a,b)$, we set $a=b=1/2$, which makes the prior on global $R^2$ be relatively diffuse.

\subsubsection*{Sparsity-inducing prior} 

If the user does not want to use the data to set the hyperparameters, $p_g>n$ for some $g$ such that the linear regression model cannot be fit and/or he wants to explicitly enforce sparsity in the regression coefficients, the following approach is also available. When there are many groups (large $G$), it is natural to prefer sparsity between groups. 
Since $R^2_g$ corresponds to the contribution of group $g$ to the total modeled variation, such a sparsity is realized by having $R^2_g \approx 0$ for several $g$'s. For this purpose, we can set $a_g<1$. If the groups are {\it a priori} indistinguishable, then we set $a_1 = \cdots = a_G$. To be coherent with the settings of the original R2D2 prior, we set $a_g = a/G$ for all $g$. 

\subsubsection*{Local variance allocation hyperparameters}

The gR2D2-D prior contains another set of Dirichlet shape parameters, $(a_{g1},\dots , a_{gp_g})$, for each $g$. To induce sparsity in regression coefficients in group $g$, one could set $a_{gj}$'s to be all equal and less than unity. In this study, for computational convenience, we use $a_{g1}=\cdots =a_{gp_g} = a_g / p_g$ as a default choice, which ensures that $\sum_{j=1}^{p_g}a_{gj}=a_g$. For details, see Section~\ref{sec:comp}. 

Finally, the gR2D2-L prior has covariance matrix $\Sigma_g$ for each $g$. We set the covariance matrix $\Sigma_g$ to approximate the Dirichlet distribution with shape parameters $(a_{g1},\dots,a_{gp_g})$. As discussed in \cite{aguilar2024generalized}, setting $\Sigma_g$ directly is not easy since the hyper-parameters lack interpretability. With this approximation, however, we inherit some interpretation from the $a_{gj}$ parameters, while still maintaining the flexibility of the logistic normal distribution. In particular, when $a_{g1}=\cdots=a_{gp_g}=1/p_g$, we set $(\Sigma_g)_{kj}=\delta(1/p_g)+I(k=j)\delta(1/p_g)$, where $\delta(x)$ is the trigamma function (second order derivatives of log-gamma function). As discussed above, small values of $a_{gj}$ help induce sparsity in the model, so the resulting $\Sigma_g$ should inherit similar properties. Please see \cite{aguilar2024generalized} for further details.

\subsection{Theoretical properties}
We now present several theoretical properties of the gR2D2 prior. For simplicity, we only consider the Dirichlet distribution on the variance allocation parameters. A desirable feature for shrinkage priors is to have high concentration at zero (to shrink parameter estimates towards zero) while also having heavy tails (to accurately estimate the signal in the data). We can quantify these two properties with the concentration of the marginal prior distribution of $\beta_{gj}$ near the origin and the asymptotical tail behavior.

\begin{prp} \label{prp:marg}
Consider the gR2D2 prior defined as 
\begin{multline*}
\beta_{gj}|\phi_{gj}, W_g\sim \mathsf{DE}\left(\phi_{gj} W_g  \right), \quad W_g|\tau^2\sim\mathsf{Gamma}(a_g,\tau^{2}), \quad \\ \tau^2\sim\mathsf{InvGamma}(b,1),
\boldsymbol{\phi}_g\sim\mathsf{Dirichlet}(a_{g1},\dots,a_{gp_g})
\end{multline*}
where $\sigma^2=1$ for simplicity.
Let $\pi(\beta_{gj})$ be the marginal distribution of $\beta_{gj}$ and $a_{g1}=\cdots=a_{gp_g}:=a_{g\pi}$. Then
$$
    \pi(\beta_{gj})
    =O\left(|\beta_{gj}|^{-(1-2a_{g\pi})}\right) \text{ as }\beta_{gj}\to0
$$
and
$$
    \pi(\beta_{gj})
    =O\left(|\beta_{gj}|^{-(2b+1)}\right) \text{ as }\beta_{gj}\to\infty.
$$
\end{prp}
This result shows that the gR2D2 prior is unbounded at the origin while also having polynomial tails, desirable features in a shrinkage prior. Moreover, as $a_{g\pi},b\to0$, the origin and tail behavior, respectively, approach $O(|\beta_{gj}|^{-1})$ which is the boundary for a proper prior, making the gR2D2 prior optimal in this sense. These results are identical to that of the original R2D2 prior \citep{zhang2022bayesian} as their marginal distributions are the same. The key difference between the gR2D2 and R2D2 priors lies not in the marginal distributions but in the dependence between coefficients. Thus, we look at the joint distribution of the coefficients within a particular group, i.e., $\bbeta_g$.

\begin{prp} \label{prp:joint}
Consider the gR2D2 prior as defined in Proposition \ref{prp:marg}. Then integrating out $W_g$ (the group variance parameter), $\bbeta_g|\psi_{gj},\phi_{gj},\tau^2\sim\mathsf{ML}({\bf 0}_{p_g},\tau^{-2}\Phi)$ has a multivariate-Laplace distribution \citep{eltoft2006multivariate, kozubowski2013multivariate} where $\Phi$ is a $p_g\times p_g$ diagonal matrix with $(\psi_{g1}\phi_{g1},\dots,\psi_{gp_g}\phi_{gp_g})$ on the diagonal.
\end{prp}

Note that a diagonal covariance for the multivariate-Laplace distribution does not imply independence, thus showing that the $\beta_{gj}$ coefficients are dependent within a group (through the influence of the $W_g$ parameter). Conditioned on $\psi_{gj}$ and $\phi_{gj}$, the joint distribution of $\bbeta_g$ is also, in fact, the same for the original R2D2 prior. It is not, however, the same if we marginalize out the effect of the variance allocation parameters. To see this, consider the first group $g=1$ and notice that for the gR2D2 prior, $\boldsymbol{\phi}_1\sim\mathsf{Dirichlet}(a_{11},\dots,a_{1p_1})^\top$, implying that $\sum_{j=1}^{p_1}\phi_{1j}=1$. On the other hand, for the R2D2 prior, $\boldsymbol{\phi}=(\phi_1,\dots,\phi_p)^\top\sim\mathsf{Dirichlet}(a_1,\dots,a_p)$. This means that if we looked only at the coefficients corresponding to group $g=1$, i.e., $(\phi_1,\dots,\phi_{p_1})$, it would not be true that $\sum_{j=1}^{p_1}\phi_j=1$. Since the variance allocation parameter distribution is different for the R2D2 and gR2D2 prior when restricted to a single group, the marginal distribution of $\bbeta_g$ will also be different.

\begin{figure}
    \centering
    \includegraphics[width=0.99\linewidth]{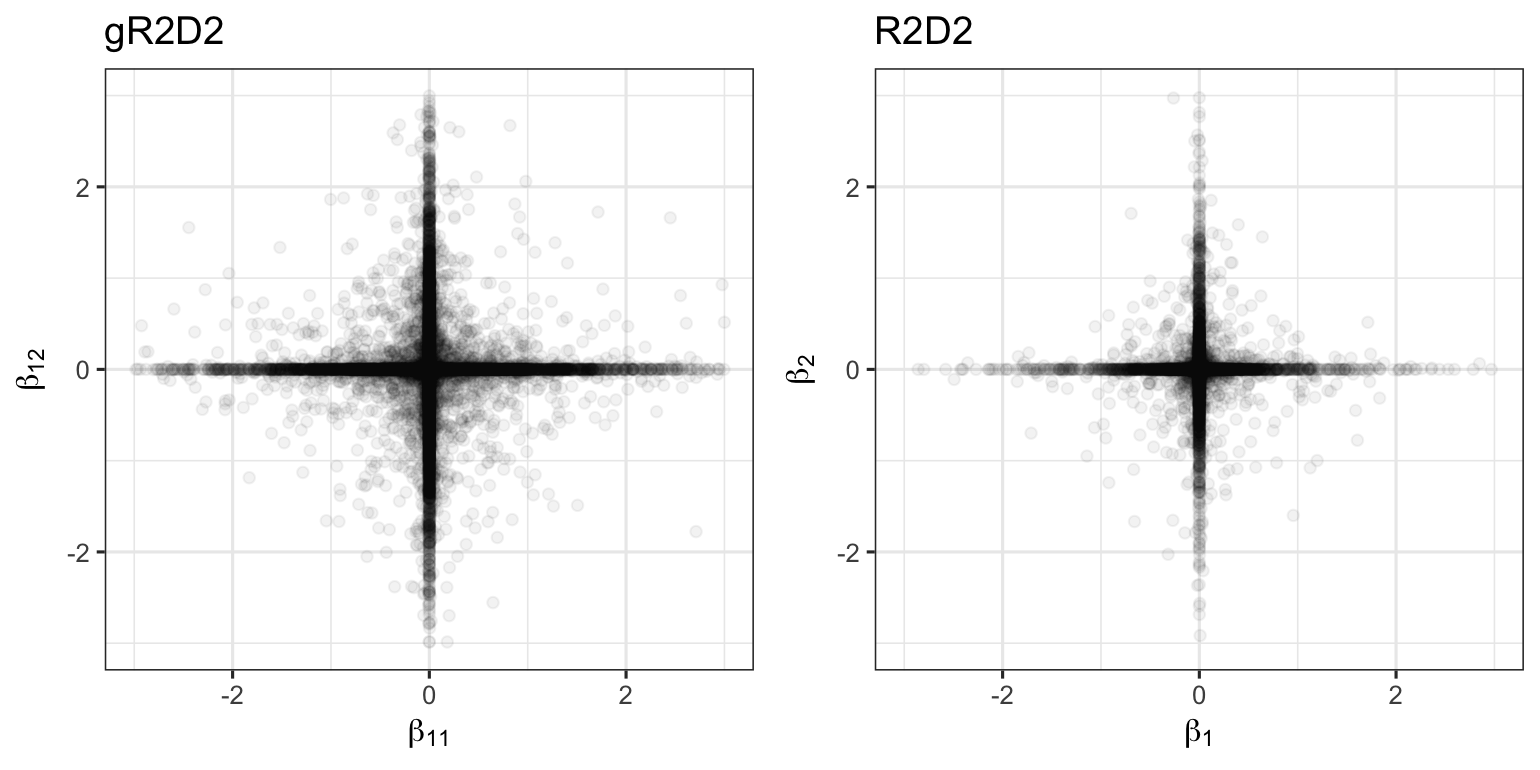}
    \caption{Realization of the joint distribution of $\bbeta_1$ for the gR2D2 and R2D2 priors.}
    \label{fig:joint}
\end{figure}

To further explore this point, we plot the marginal distribution of $\bbeta_g$ for the R2D2 and gR2D2 priors. For simplicity, let $G=2$, $p=4$ and $p_1=p_2=2$. We also set $\psi_{gj}=1$ for all $g,j$ and $\tau^2=1$ for convenience. 
For the gR2D2 prior, we first sample $\boldsymbol{\phi}_1=(\phi_{11},\phi_{12})^\top\sim\mathsf{Dirichlet}(a_{11},a_{12}$) where $a_{11}=a_{12}=a_1/p_1$ with $a_1=a/G$ and $a=1/2$. Then conditioned on this sample, we make a draw from the multivariate-Laplace distribution described in Proposition \ref{prp:joint} to obtain a sample of $\bbeta_1=(\beta_{11},\beta_{12})^\top$. We follow a similar process to make draws from the R2D2 prior, except now we draw $\boldsymbol{\phi}=(\phi_1,\dots,\phi_p)^\top\sim\mathsf{Dirichlet}(a_1,\dots,a_p)$ where $a_1=\cdots=a_p:=a/p$ and keep only $\tilde{\boldsymbol{\phi}}=(\phi_1,\phi_2)^\top$. Again conditioned on this draw, we sample $\bbeta_1=(\beta_{11},\beta_{12})^\top$. We plot the results of $10,000$ draws in Figure \ref{fig:joint} to visualize the difference in the joint distribution induced by the gR2D2 prior compared to the original R2D2 prior. Indeed, the former yields a much more ``diffuse'' distribution while the latter is more concentrated around the origin. This result is sensible as the full Dirichlet distribution of the gR2D2 prior will have a greater variance than simply taking the first two components of a larger Dirichlet distribution in the R2D2 case. The results are similar for other parameter values and thus not included here.

\subsection{Discussion on gR2D2 prior}
We now make several observations about the gR2D2 prior. 
First, we note that we defined the group-specific $R^2_g$ as $R^2_g=\mathsf{Var}(\bx_g^{\top}\bbeta_g)/\{\mathsf{Var}(\bx^{\top}\bbeta)+\sigma^2\}$ where the denominator takes the variance over all $\bbeta$. Another valid definition is $\tilde R^2_g=\mathsf{Var}(\bx_g^{\top}\bbeta_g)/\{\mathsf{Var}(\bx_g^{\top}\bbeta_g)+\sigma^2\}=W_g/(W_g+1)$ where now the numerator and denominator variances are only over the group-specific coefficients $\bbeta_g$. Indeed, this definition induces the same prior distribution on $W_g$. We prefer the first definition, however, as it gives $\sum_{g=1}^G R_g^2=R^2$, and has a more natural interpretation as the proportion of total variance in the response explained by the coefficients in group $g$. In the second definition, $R^2_g$ is the amount of variation in group $g$ explained by $\bbeta_g$. As noted above, $R^2_g$ also serves as a relative measure of the importance of group $g$ to the overall model, something $\tilde R^2_g$ does not. The preferred definition also leads to interpretable hyperparameters. Under fixed $b$, the values of $a_1,\dots,a_G$ control the amount of group-specific $R_g^2$, and their sum controls the overall $R^2$, i.e., large values of $a_g$ favor large $R^2_g$ and large $\sum_{g=1}^G a_g$ favor large $R^2$. On the other hand, $b$ controls the contribution of the error term to the total variation of the response. We also recover the results of \cite{zhang2022bayesian} by setting $G=1$ in the proposed prior, meaning that the proposed prior is a proper generalization of the original R2D2 prior.

\subsection{Comparison with GIGG prior}
We can also compare our prior distribution with the GIGG prior \citep{boss2023group}. The GIGG prior is formulated as
\begin{align*}
&\beta_{gj}|\tau^2,\gamma_g^2, \lambda_{gj}^2\sim\mathsf{Normal}(0,\tau^2\gamma_g^2\lambda_{gj}^2),\\
\tau\sim\mathsf{C}^+&(0,\sigma),\ \gamma_g^2\sim\mathsf{Gamma}(a_g,1),\ \lambda_{gj}^2\sim\mathsf{InvGamma}(b_g,1).
\end{align*}
The connections with our model are now clear. The  gR2D2 global variance parameter ($W$) has a Beta-prime distribution, while the global parameter in GIGG $(\tau^2)$ is a squared half-Cauchy distribution. Moreover, the global variance parameter is the sum of the grouped variance parameters in gR2D2, whereas the global and group variance parameters are independent in GIGG. As for the group-variance prior distribution, the prior for $W_g$ conditional on $\tau^2$ follows a gamma distribution; the comparable parameter in GIGG $(\gamma_g^2)$ is unconditionally gamma distributed. Again, the $W_g$'s are correlated via $\tau^2$ for gR2D2, but the $\gamma_g^2$ are independent in GIGG. Lastly, the local variance allocation is controlled with either a Dirichlet or logistic-normal distribution in gR2D2, but is endowed with independent Beta-prime distributions in GIGG. Thus, at each level of the model, the gR2D2 prior encodes dependence among the variance parameters, while all such parameters are independent for GIGG. While this dependence leads to good empirical performance, it make deriving further theoretical results, e.g., posterior consistency, more difficult, so we postpone this for future work.

We may also compare the marginal properties of $\beta_{gj}$ between the gR2D2 and GIGG priors. First, for the asymptotic tail behavior, we showed in Proposition \ref{prp:marg} that the gR2D2 prior behaves like $O(|\beta_{gj}|^{-(2b+1)})$. In Theorem 2.1 of \cite{boss2023group}, the authors show that the GIGG prior is $O(|\beta_{gj}|^{-(2b_g+1)})$ for hyper-parameter $b_g$. Thus, the two priors exhibit similar (marginal) tail behavior. For concentration around the origin, we found that the gR2D2 prior was $O(|\beta_{gj}|^{-(1-2a_{g\pi})})$. Unfortunately, neither \cite{boss2023group} nor \cite{bai2019large} precisely characterize the origin behavior, only stating that it is unbounded at 0. We note, however, that the marginal prior distribution of the GIGG prior is the same as that of the gR2D2 prior (for $\sigma^2=1$), except that the former uses a normal distribution for $\beta_{gj}$ while the latter uses a Laplace distribution. Since the Laplace prior has greater concentration at 0, we expect that the gR2D2 prior will have more prior mass at 0 compared with the GIGG prior. We see evidence of this in both the simulation study and real-data analysis.

Finally, the parameter values themselves are also less interpretable in the GIGG model, in particular $\tau^2$. Ideally, the value of $\tau^2$ should relate to the overall variance of the model, e.g., a small $\tau^2$ means that there is very little signal in the data. Similarly, $\gamma_g^2>\gamma_{g'}^2$ should encode that group $g$ has more variability than group $g'$. We claim that this latter condition is true but the former is not. Indeed, if we scale by a constant $c$, yielding $\tau^2/c$ and $c\gamma_g^2$, then their product is the same as is their relative comparison of $c\gamma_g^2,c\gamma_{g'}^2$, but the value of $\tau^2/c$ (global variance) changes. Therefore, it is not appropriate to consider $\tau^2$ as the global variance in this sense. On the other hand, in the gR2D2 model, due to the constraints introduced in Proposition \ref{prp:r2d2}, $W=\sum_gW_g$ is a proper global variance parameter and $W_g$ (or more precisely $R^2_g$) yields the relative sense of group variance.

\section{Posterior Computation}\label{sec:comp}

\subsection{Joint posterior and full conditional distributions for gR2D2 priors}

We now present the full MCMC sampling scheme for the gR2D2 prior. Let $\bm{Y}=(Y_1,\dots,Y_n)^{\top}$, $\bm{X}=(\bm{x}_1 ,\cdots ,\bm{x}_n)^\top$, and $\boldsymbol{W}$, $\Psi$ and $\Phi$ be collections of $W_g$, $\psi_{gj}$ and $\phi_{gj}$, respectively. 
We define $\mathsf{InvGauss}( \mu, \lambda )$ as the inverse-Gaussian distribution with density $p(z;\mu,\lambda)\propto z^{-3/2}\exp\left\{-\lambda(z-\mu)^2/(2\mu^2z)\right\}$ and $\mathsf{GIG}(\lambda,\rho,\chi)$ as the generalized inverse-Gaussian with density $p(z|\lambda,\rho,\chi)\propto z^{\lambda-1}\exp\left\{-(\rho z+\chi/z)/2\right\}$.

%
For the error variance, we assign an inverse-gamma prior, $\sigma^2\sim \mathsf{InvGamma}(n_0/2, d_0/2)$.
The joint posterior of $\bbeta$ and latent variables is given by 
\begin{align*}
\pi(\bbeta,\sigma^2 ,&\tau^2, \Psi,\Phi,\boldsymbol{W} |\bm{Y})\\
&\propto \exp\left\{-\frac1{2\sigma^2}(\bm{Y}-\bm{X}\bbeta)^\top (\bm{Y}-\bm{X}\bbeta)
- \frac{1}{2\sigma^2}\sum_{g=1}^G\sum_{j=1}^{p_g}\frac{\beta_{gj}^2}{ \psi_{gj} \phi _{gj}  W_g /2}
\right\}\\
&\times \exp\Big(-\frac{d_0}{2\sigma^2} -\frac{1}{\tau^2}\Big)
(\sigma^2)^{-(n_0+n+p)/2-1} (\tau^2)^{-\sum_{g=1}^G a_g-b-1}  
\prod_{g=1}^G \pi(\bm{\phi}_g) \\
&\times \prod_{g=1}^G  W_g^{a_g-p_g/2-1}e^{-W_g/\tau^2}\prod_{j=1}^{p_g} (\phi_{gj}\psi_{gj})^{-1/2} e^{-\psi_{gj}/2},
\end{align*}
where $\pi(\bm{\phi}_g)$ is the prior distribution of $\bm{\phi}_g$, being either Dirichlet or logistic-normal densities. Targeting this joint posterior distribution, we propose the following Gibbs sampler algorithm.

\begin{itemize}
\item[-] \textbf{Step 1:} sampling from $\pi(\bbeta|\sigma^2 ,\tau^2,\Psi,\Phi,\boldsymbol{W}, \bm{Y})$.

The full conditional distribution of $\boldsymbol{\beta}$ is $\mathsf{Normal}(\mu _{\beta}, \sigma^2 \Sigma_{\beta})$, where
\begin{equation*}
\Sigma_{\beta} = ( \bm{X}^{\top}\bm{X} + \bm{\Lambda}^{-1} )^{-1}, \ \ \ \ \  \ \ \ \mu_{\beta} = ( \bm{X}^{\top}\bm{X} + \bm{\Lambda} ^{-1} )^{-1} \bm{X}^{\top} \bm{Y},
\end{equation*}
and $\bm{\Lambda}$ is the diagonal matrix whose $(j + \sum _{g'=1}^{g-1} p_{g'})$-th diagonal element is $\psi_{gj} \phi_{gj} W_g/2$.
	
\item[-] \textbf{Step 2:} sampling from $\pi(\sigma^2 |\bbeta, \tau^2,\Psi,\Phi,\boldsymbol{W}, \bm{Y})$.

The full conditional distribution of $\sigma^2$ is $\mathsf{InvGamma}( n_1/2, d_1/2 )$, where 
\begin{equation*}
n_1 = n_0 + n + p, \ \ \ \ \  \ d_1 = d_0 + ( \bm{Y} - \bm{X}\boldsymbol{\beta} )^{\top} ( \bm{Y} - \bm{X}\boldsymbol{\beta} ) + \boldsymbol{\beta}^{\top} \bm{\Lambda}^{-1} \boldsymbol{\beta}.
\end{equation*}

\item[-] \textbf{Step 3:} sampling from $\pi(\tau ^2 | \bbeta, \sigma^2 , \Psi,\Phi,\boldsymbol{W}, \bm{Y})$.

The full conditional distribution of $\tau^2$ is $\mathsf{InvGamma}( a_{\tau}, b_{\tau} )$, where 
\begin{equation*}
a_{\tau} = b + \sum_{g=1}^G a_g \ \ \ \ \ b_{\tau} = 1 + \sum_{g=1}^G W_g.
\end{equation*}

\item[-] \textbf{Step 4:} sampling from $\pi(\Psi|\bbeta, \sigma^2 ,\tau ^2 , \Phi,\boldsymbol{W}, \bm{Y})$.

Conditional on the other variables, $\psi_{gj}$'s are mutually independent. The full conditional distribution of $\psi_{gj}^{-1}$ is $\mathsf{InvGauss}( \mu _{\psi, gj}, \lambda_{\psi} )$, where
\begin{equation*}
\mu _{\psi, gj} = \frac{ \sqrt{ \sigma^2 \phi _{gj} W_g /2 }  }{ |\beta _{gj}| }, \ \ \ \ \ \ \ \lambda_{\psi} = 1.
\end{equation*}

\item[-] \textbf{Step 5:} sampling from $\pi(\Phi,\boldsymbol{W}| \bbeta, \sigma^2 ,\tau ^2,\Psi ,  \bm{Y})$.

We first notice the conditional independence:
\begin{equation*}
    \pi(\Phi,\boldsymbol{W}| \bbeta, \sigma^2 ,\tau ^2 , \Psi, \bm{Y}) = \prod _{g=1}^G \pi(\boldsymbol\phi_g,W_g| \bbeta, \sigma^2 ,\tau ^2 , \Phi, \bm{Y}).
\end{equation*}
Thus we can sample from $\pi(\boldsymbol\phi_g,W_g| \bbeta, \sigma^2 ,\tau ^2 , \Psi, \bm{Y})$ in parallel. This step depends on the choice of priors and is discussed in more detail next, as it is the only step of the Gibbs sampler which does not yield familiar forms. 
\end{itemize}

We note that sampling of $\bm{\beta}$ in Step 1 requires the Cholesky decomposition of a $p{\times}p$ matrix at every iteration and could be computationally demanding when $p$ is larger than $n$. For this issue, we additionally utilize the scalable sampling strategy proposed in \cite{bhattacharya2016fast} in Step~1.

\subsection{Gibbs sampler for gR2D2-D}

Sampling of $(\boldsymbol\phi_g,W_g)$ is the two-step program: sampling $\boldsymbol\phi_g$ from the marginal, then sampling $W_g$ conditional on $\boldsymbol\phi_g$. Those conditional posteriors are not well-known distributions in general, but greatly simplified when $a_g = \sum _{j=1}^{p_g}a_{gj}$. We start from this special case, then move on to the general choice of shape parameters. 

In the gR2D2-D model, we assume $\boldsymbol{\phi}_g \sim \mathsf{Dirichlet} (a_{g1},\dots ,a_{gp_g})$. For each $g$, let $a_g^{\ast} = \sum _{j=1}^{p_g} a_{gj}$. 
In the case that $a_g = a_g^{\ast}$, the exact sampling from the distributions at Step~5 is feasible. The algorithm below is proposed by \cite{zhang2022bayesian}, motivated by the method used for the Dirichlet Laplace model \citep{bhattacharya2015dirichlet}. 
\begin{itemize}

\item[-] \textbf{Step 5a:} Sampling from $\pi(\boldsymbol\phi_g| \bbeta, \sigma^2 ,\tau ^2 , \Psi ,\bm{Y})$.

The marginal distribution of $\boldsymbol\phi_g$ satisfies the following distributional equation: 
\begin{equation}\label{eq:phi-pos}
\phi_{gj} = \frac{T_{gj}}{ T_{g,1} + \cdots + T_{g,p_g} }, 
\end{equation}
where $T_{g,1}, \dots, T_{g,p_g}$ are independent and distributed as 
\begin{equation*}
T_{gj} \sim \mathsf{GIG}\left( a_{gj} -\frac{1}{2} , 2, \frac{\beta_{gj}^2}{ \sigma^2\psi_{gj}W_g/2 } \right).
\end{equation*}

\item[-] \textbf{Step 5b:} Sampling from $\pi(W_g| \bbeta, \sigma^2 ,\tau ^2 , \boldsymbol\phi_g, \Psi, \bm{Y})$.

Conditional on the sampled $\boldsymbol\phi_g$, we sample $W_g$ as: 
\begin{equation*}
W_g \sim \mathsf{GIG}\left( a_g-\frac{p_g}{2}, \ \frac{2}{\tau^2},\  \sum_{j=1}^{p_g} \frac{\beta_{gj}^2}{ \sigma^2\psi_{gj}\phi_{gj}/2} \right).
\end{equation*}

\end{itemize}

In general, we have $a_g \not= a_g^{\ast} = \sum _{j=1}^{p_g} a_{gj}$ and cannot use the algorithm above as-is. In fact, the sampling steps above are valid when we use $W_g \sim \mathsf{Gamma}(a_g^{\ast},1/\tau^2 )$, instead of $W_g \sim \mathsf{Gamma}(a_g,1/\tau^2)$. Based on this observation, we propose to use the full conditional posterior under the prior with $a_g^{\ast}$ as the proposal distribution, then add the acceptance-rejection step to justify it as the Metropolis-Hastings algorithm. The acceptance probability is then simply the ratio of gamma densities. 

\begin{itemize}
\item[-] Let $( \boldsymbol\phi _g^{\rm old} , W_g^{\rm old} )$ be a sample generated at the previous iteration. 

\item[-] \textbf{Steps 5a and 5b (gR2D2-D):} Set $a_g^{\ast}=\sum _{j=1}^{p_g} a_{gj}$ and sample $( \boldsymbol\phi _g^{\rm new} , W_g^{\rm new} )$ by following the step 5a and 5b described before. 

\item[-] \textbf{Step 5c (gR2D2-D):} Accept $( \boldsymbol\phi _g^{\rm new} , W_g^{\rm new} )$ as the posterior sample of the current iteration with probability 
\begin{equation*}
    \min \left\{ 1, \left( \frac{W_g^{\rm new}}{ W_g^{\rm old} } \right) ^{a_g^{\ast}-a_g} \right\},
\end{equation*}
otherwise accept $( \boldsymbol\phi _g^{\rm old} , W_g^{\rm old} )$.
\end{itemize}

In practice, the acceptance rate needs to be monitored to ensure efficiency of the algorithm. 

\subsection{Gibbs sampler for gR2D2-L} \label{sec:sampL}

In the gR2D2-L model, we assume $\boldsymbol\phi_g\sim\mathsf{LogisticNormal}(\mathbf{0}_{p_g-1},\Sigma_g)$. Even under this prior, we can take a similar sampling approach as in the previous sub-section. The challenge, however, is that the logistic-normal prior lacks shape parameters $(a_{g1},\dots g_{gp_g})$ to use in the proposal distribution. To overcome this, we simply set $a_{gj} = a_g/p_g$ so that $a_g = a_g^{\ast}$ holds. The difference of the priors for $\boldsymbol\phi_g$ in the target and proposal distributions modifies the acceptance probability of the step 5c.

\begin{itemize}
\item[-] \textbf{Steps 5a and 5b (gR2D2-L):} Set $a_{gj}=a_g/p_g$ and sample $( \boldsymbol\phi _g^{\rm new} , W_g^{\rm new} )$ by following the step 5a and 5b described before. 

\item[-] \textbf{Step 5c (gR2D2-L):} Accept $( \boldsymbol\phi _g^{\rm new} , W_g^{\rm new} )$ as the posterior sample of the current iteration with probability $\min \{ 1 , f(\boldsymbol\phi_g^{\rm new}) / f(\boldsymbol\phi_g^{\rm old}) \}$, where $f(\cdot )$ is given by, 
$$
\log f(\boldsymbol{\phi})\equiv -\sum_{j=1}^{p_g} a_{gj} \log \phi_{gj} - \frac12 {\rm LR}(\boldsymbol{\phi}_g)^\top \Sigma_g^{-1}{\rm LR}(\boldsymbol{\phi}_g).
$$
\end{itemize}

\section{Simulations}\label{sec:sim}
We now demonstrate the performance of the gR2D2 prior on simulated data by looking at the estimation accuracy and inference properties. All code to implement the proposed and competing methods is available on GitHub: \url{https://github.com/sshonosuke/GR2D2/tree/main}.

\subsection{Settings}
We consider three cases for the sample size and the number of covariates, i.e., $(n,p)\in\{(250, 50), (150, 200), (200, 1000)\}$.
The coefficients are grouped into $G=5$ (when $p=50$), $G=20$ ($p=200$) and $G=100$ ($p=1000$) groups of $p_g=10$ coefficients for $j=1,\dots,G$.
We generate covariates $\bx_i\sim\mathsf{Normal}({\bf 0}_p,\Sigma)$ where $\Sigma$ has a block correlation structure, i.e., $\Sigma_{ij}=0.7$ if coefficients $i$ and $j$ are in the same group, and $0.2$ otherwise. 
We consider the following five scenarios for signal (non-null) values of $\beta$: 
\begin{align*}
&{\rm (s1)} \ \ \beta_{11},\beta_{21},\beta_{41}=5, \ \ \beta_{31},\beta_{51}=-5\\
&{\rm (s2)} \ \ \beta_{11},\beta_{12}, \beta_{13}=3,\ \ \beta_{21}, \beta_{22},\beta_{23}=-3, \ \ \beta_{31}=2, \ \ \beta_{51},\beta_{52},\beta_{53},\beta_{54},\beta_{55}=-1\\
&{\rm (s3)} \ \ \beta_{11},\beta_{12}, \beta_{13}=3,\ \ \beta_{21}, \beta_{22},\beta_{23}=-3, \ \ \beta_{51}=3, \beta_{55}=4 \\
&{\rm (s4)} \ \ \beta_{11}, \beta_{13},\beta_{15},\beta_{17},\beta_{19}=3, \ \ \beta_{31},\beta_{32},\beta_{33},\beta_{34},\beta_{35}=-2\\
&{\rm (s5)} \ \ \beta_{11}, \beta_{15},\beta_{19}=5, \ \ \beta_{13},\beta_{17}=-5
\end{align*}
Note that $\sum_{g=1}^G\sum_{j=1}^{p_g}|\beta_{gj}|=25$ for all the scenarios, and number of null groups increases from 0 to 4 going from scenario 1 to 5. In scenario 1, there is one significant coefficient in each group, while four groups have significant coefficients in scenario 2, with a differing number of signals in each group. Significant coefficients are roughly equally distributed between three and two groups in settings 3 and 4, respectively, whereas all non-null coefficients are in a single group in setting 5. We set $\sigma^2$ so that the resulting signal-to-noise ratio (SNR) matches a pre-specified value (0.5 and 0.7).

We compare the gR2D2 prior with several other methods. As a baseline, we consider prior frameworks that do not account for grouping structure, including the original R2D2 prior \citep{zhang2022bayesian} and the horseshoe prior (HS) \citep{carvalho2009handling}. 
We also consider shrinkage priors designed for the grouped setting including the GIGG prior \citep{boss2023group} and grouped horseshoe prior (GHS) \citep{xu2016bayesian}. We wrote custom Gibbs samplers for posterior computation for all competing methods.  
Finally, for the gR2D2 prior, the variance allocation parameters follow either the Dirichlet distribution (gR2D2-D) or the logistic normal distribution (gR2D2-L).
For both methods, we set the hyperparameters as in Section~\ref{sec:hyper}, using the Empirical Bayes approach for gR2D2-D.


%
For all methods, we generate 1000 posterior samples of $\bbeta$ after discarding the first 1000 samples as burn-in, and then compute the posterior mean and $95\%$ credible intervals.
Our metric of interest is the sum of squared errors between the posterior mean $\hat{\bbeta}$ and the true value $\bbeta$, averaged over the number of simulations ($R=100$):
$$
{\rm MSE}_{gj}
=\frac1{R}\sum_{r=1}^R (\widehat{\beta}^{(r)}_{gj}-\beta_{gj})^2, \ \ \ \ k=1,\ldots,p,
$$
where $\widehat{\beta}^{(r)}_{gj}$ denotes the posterior mean of $\beta_{gj}$ in the $r$th replication. 
The MSE values are aggregated over $g=1,\ldots,G$ and $j=1,\ldots,p_g$, but separated between null and non-null coefficients.
Furthermore, we evaluate $95\%$ credible intervals ${\rm CI}_{gj}^{(r)}$ of regression coefficients by empirical coverage probability (CP), and average lengths (AL) of the credible interval $|{\rm CI}_{gj}^{(r)}|$, defined as 
$$
{\rm CP}_{gj}=\frac1R\sum_{r=1}^R I(\beta_{gj}\in {\rm CI}_{gj}^{(r)}), \ \ \ \ \ \ 
{\rm AL}_{gj}=\frac1R\sum_{r=1}^R |{\rm CI}_{gj}^{(r)}|.
$$
The values of CP and AL are averaged over $g=1,\ldots,G$ and $j=1,\ldots,p_g$.

\subsection{Results}
The MSE results are in Table~\ref{tab:sim}. Note that the acceptance rates of the MH step in gR2D2-L (Step 5c (gR2D2-L) in Section~\ref{sec:sampL}) were around $84\%$ in all the scenarios. For $n=250$ and $p=50$, the gR2D2-D prior provides the most accurate estimates for null-signals, while GIGG, GHS and gR2D2-D  perform well in estimating signals. In particular, in scenario 1 where each group is the same with only a single non-null coefficient, incorporating the group structure is not essential in improving the model fit, resulting in the non-grouping methods like R2D2 and HS performing as well or better than other group-based approaches. In scenario 5, on the other hand, modeling the between-group sparsity is the most important as all the signals are contained in a single group. By setting the hyperparameters of the gR2D2-D prior using the Empirical Bayes approach, this prior performs particularly well in this scenario. 

When the signal is distributed between several, but not all groups as in scenarios 2-4, the GIGG prior yields the best estimates of the non-null coefficients, particularly when SNR $=0.5$. For larger SNR ($0.7$) and Scenarios 3 and 5, the performance of the gR2D2-D prior improves and outperforms the GIGG prior. In terms of the choice of the prior for the variance allocation parameters, we see that the Dirichlet distribution (gR2D2-D) yields better estimation of the null coefficients, while the logistic normal distribution (gR2D2-L) tends to provide lower MSE values on the non-null coefficients. Lastly, both gR2D2-D and gR2D2-L generally outperform the original R2D2 prior, exemplifying the importance of incorporating the group-structure into the modeling framework. The results are similar when $n=150$ and $p=200$. 

In the highest dimensional setting where $n=200$ and $p=1000$, the gR2D2-D prior demonstrates strong performance. In particular, it generally has the smallest or second smallest estimation error for both null and non-null coefficients. Moreover, its estimation accuracy is often an order of magnitude smaller than that of some competing methods, i.e., null coefficients in Scenario 1 and non-null coefficients in Scenario 5.

We next discuss the model performance in posterior uncertainty quantification by the 95\% posterior credible intervals, reported in Table~\ref{tab:sim-cp}. 
We notice the over-coverage in all methods, settings and scenarios. Given all the methods can achieve the nominal 95\% coverage, we prefer methods with shorter credible intervals. In each setting, gR2D2-D provides the narrowest intervals, while R2D2 and gR2D2-L always yield the next narrowest intervals. The credible intervals of the gR2D2-L prior are slightly wider, likely due to the extra parameters and model flexibility. Both of the proposed gR2D2 methods have narrower intervals than those of GIGG for both SNRs. This is likely because we believe the gR2D2 prior distribution has a higher concentration of prior mass at 0 compared with that of GIGG.


\begin{table}[htb!]
\centering
{\small 
\begin{tabular}{ccccccccccccccc}
\hline
\multicolumn{3}{c}{$(n,p)=(250, 50)$}& \multicolumn{5}{c}{Null} &&  \multicolumn{5}{c}{Non-null}\\
SNR & Scenario &  & 1 & 2 & 3 & 4 & 5 &  & 1 & 2 & 3 & 4 & 5 \\
\hline
 & R2D2 &  & 1.67 & 1.64 & 2.40 & 3.25 & 0.77 &  & {\bf 3.63} & 13.8 & 14.1 & 31.7 & 3.99 \\
 & HS &  & 1.86 & 2.36 & 3.20 & 3.67 & 0.86 &  & 4.32 & 12.0 & 13.5 & 27.0 & 4.96 \\
0.5 & GIGG &  & 10.4 & 8.06 & 10.5 & 11.9 & 6.11 &  & 4.84 & {\bf 8.65} & {\bf 9.85} & {\bf 16.3} & 5.02 \\
 & GHS &  & 2.41 & 3.06 & 3.97 & 4.18 & 1.93 &  & 4.56 & 11.7 & 11.9 & 19.0 & 4.45 \\
 & gR2D2-D &  & {\bf 1.11} & {\bf 1.00} & {\bf 1.48} & {\bf 1.62} & {\bf 0.18} &  & 4.29 & 16.8 & 14.5 & 30.9 & {\bf 3.93} \\
 & gR2D2-L &  & 1.68 & 1.50 & 2.12 & 2.63 & 0.53 &  & 3.94 & 13.6 & 13.6 & 29.6 & 4.11 \\
\hline
 & R2D2 &  & 0.65 & 0.57 & 0.71 & 1.31 & 0.39 &  & {\bf 1.31} & 4.94 & 3.51 & 12.4 & 1.46 \\
 & HS &  & 0.73 & 1.21 & 1.35 & 2.22 & 0.41 &  & 1.39 & 4.41 & 4.00 & 11.1 & 1.57 \\
0.7 & GIGG &  & 5.66 & 4.43 & 5.64 & 7.00 & 3.37 &  & 1.70 & {\bf 3.66} & 3.83 & {\bf 7.84} & 1.77 \\
 & GHS &  & 0.94 & 1.71 & 1.96 & 2.61 & 0.96 &  & 1.42 & 4.36 & 3.82 & 7.98 & 1.62 \\
 & gR2D2-D &  & {\bf 0.47} & {\bf 0.37} & {\bf 0.41} & {\bf 0.61} & {\bf 0.10} &  & 1.36 & 6.33 & {\bf 3.46} & 11.8 & {\bf 1.44} \\
 & gR2D2-L &  & 0.67 & 0.54 & 0.65 & 1.02 & 0.26 &  & 1.37 & 4.86 & 3.54 & 11.6 & 1.48 \\
\hline
\end{tabular}
\medskip 
\begin{tabular}{ccccccccccccccc}
\hline
\multicolumn{3}{c}{$(n,p)=(150, 200)$}& \multicolumn{5}{c}{Null} &&  \multicolumn{5}{c}{Non-null}\\
SNR & Scenario &  & 1 & 2 & 3 & 4 & 5 &  & 1 & 2 & 3 & 4 & 5 \\
\hline
 & R2D2 &  & 28.5 & 32.4 & 38.7 & 37.4 & 11.8 &  & {\bf 15.7} & 44.0 & 41.9 & 48.2 & 12.4 \\
 & HS &  & 3.50 & 6.52 & 6.22 & 5.04 & 1.15 &  & 19.3 & 50.2 & 50.8 & 58.3 & 28.5 \\
0.5 & GIGG &  & 43.5 & 43.9 & 48.8 & 47.8 & 28.8 &  & 22.1 & {\bf 30.8} & {\bf 31.6} & {\bf 31.6} & 19.4 \\
 & GHS &  & 10.3 & 10.7 & 8.78 & 7.13 & 6.27 &  & 19.7 & 36.2 & 36.7 & 35.7 & {\bf 9.49} \\
 & gR2D2-D &  & {\bf 2.20} & {\bf 5.10} & {\bf 5.55} & {\bf 4.46} & {\bf 0.40} &  & 26.4 & 42.5 & 45.1 & 49.2 & 10.6 \\
 & gR2D2-L &  & 13.2 & 15.8 & 18.2 & 15.7 & 5.17 &  & {\bf 15.7} & 39.4 & 37.9 & 43.9 & 11.1 \\
 \hline
 & R2D2 &  & 12.1 & 13.8 & 15.2 & 19.0 & 5.67 &  & 3.34 & 22.9 & 21.6 & 30.6 & 3.29 \\
 & HS &  & 1.41 & 2.54 & 3.28 & 2.98 & 0.42 &  & {\bf 2.72} & 28.3 & 26.4 & 38.0 & 3.26 \\
 & GIGG &  & 28.9 & 29.2 & 32.0 & 35.0 & 19.3 &  & 6.28 & {\bf 18.8} & 18.4 & 22.4 & 6.38 \\
0.7 & GHS &  & 2.71 & 4.23 & 5.35 & 4.21 & 1.73 &  & 3.11 & 19.7 & {\bf 16.9} & {\bf 18.8} & 3.20 \\
 & gR2D2-D &  & {\bf 1.26} & {\bf 2.12} & {\bf 2.15} & {\bf 1.90} & {\bf 0.15} &  & 2.78 & 24.6 & 19.4 & 27.4 & {\bf 2.73} \\
 & gR2D2-L &  & 6.15 & 6.46 & 7.64 & 7.74 & 2.13 &  & 3.18 & 20.7 & 18.8 & 25.9 & 3.10 \\
\hline
\end{tabular}
\medskip 
\begin{tabular}{ccccccccccccccc}
\hline
\multicolumn{3}{c}{$(n,p)=(200, 1000)$}& \multicolumn{5}{c}{Null} &&  \multicolumn{5}{c}{Non-null}\\
SNR & Scenario &  & 1 & 2 & 3 & 4 & 5 &  & 1 & 2 & 3 & 4 & 5 \\
\hline
 & R2D2 &  & 35.6 & 40.8 & 54.8 & 71.0 & 16.9 &  & 25.2 & 48.7 & 59.9 & 66.5 & 15.6 \\
 & HS &  & 16.3 & 25.2 & 37.5 & 54.0 & 2.76 &  & 22.3 & 49.9 & 57.8 & 66.3 & 32.6 \\
0.5 & GIGG &  & 9.39 & 10.1 & 12.6 & 16.9 & 5.32 &  & 62.2 & 38.9 & 49.0 & 42.4 & 74.4 \\
 & GHS &  & 40.0 & 45.7 & 55.3 & 67.4 & 23.9 &  & 21.6 & {\bf 33.2} & {\bf 41.4} & {\bf 40.4} & 7.76 \\
 & gR2D2-D &  & {\bf 3.54} & {\bf 6.29} & {\bf 7.35} & {\bf 8.79} & {\bf 1.08} &  & {\bf 11.5} & 36.2 & 42.2 & 49.4 & {\bf 5.00} \\
 & gR2D2-L &  & 65.1 & 72.4 & 92.7 & 123 & 32.9 &  & 23.0 & 40.5 & 49.2 & 56.6 & 9.37 \\
 \hline
 & R2D2 &  & 12.3 & 15.9 & 19.1 & 27.5 & 6.26 &  & 3.65 & 30.2 & 31.2 & 47.5 & 2.61 \\
 & HS &  & {\bf 0.93} & 4.33 & 5.72 & 11.3 & {\bf 0.05} &  & {\bf 2.09} & 32.9 & 36.8 & 49.0 & 3.72 \\
 & GIGG &  & 4.27 & 4.76 & 5.69 & 7.04 & 2.86 &  & 34.0 & 31.7 & 38.9 & 34.0 & 57.8 \\
0.7 & GHS &  & 17.6 & 22.3 & 24.0 & 33.6 & 11.0 &  & 4.09 & 19.1 & 18.7 & {\bf 23.2} & 2.98 \\
 & gR2D2-D &  & 1.41 & {\bf 2.63} & {\bf 2.44} & {\bf 2.98} & 0.47 &  & 2.61 & {\bf 18.8} & {\bf 16.8} & 30.1 & {\bf 2.01} \\
 & gR2D2-L &  & 26.2 & 30.7 & 34.5 & 53.8 & 12.4 &  & 4.66 & 23.5 & 23.6 & 36.5 & 3.00 \\
\hline
\end{tabular}
}
\caption{MSE under five scenarios of regression coefficients and two values of SNR. The lowest value for each setting is represented in {\bf bold}.
}
\label{tab:sim}
\end{table}

\begin{table}[htb!]
\centering
{\small
\begin{tabular}{ccccccccccccccc}
\hline
\multicolumn{3}{c}{$(n,p)=(250, 50)$}& \multicolumn{5}{c}{SNR$=0.5$} && \multicolumn{5}{c}{SNR$=0.7$}\\
 & Scenario &  & 1 & 2 & 3 & 4 & 5 &  & 1 & 2 & 3 & 4 & 5 \\
 \hline
 & R2D2 &  & 99.3 & 96.4 & 98.1 & 96.8 & 99.3 &  & 99.4 & 97.5 & 98.9 & 97.9 & 99.4 \\
 & HS &  & 99.2 & 96.7 & 97.7 & 96.8 & 99.0 &  & 99.4 & 97.5 & 98.5 & 97.6 & 99.3 \\
CP & GIGG &  & 98.2 & 97.1 & 97.2 & 97.5 & 97.9 &  & 98.0 & 96.9 & 97.3 & 97.2 & 97.6 \\
 & GHS &  & 99.1 & 96.4 & 97.9 & 97.6 & 99.1 &  & 99.4 & 97.5 & 98.4 & 98.1 & 99.2 \\
 & gR2D2-D &  & 99.2 & 92.9 & 98.0 & 96.7 & 99.4 &  & 99.4 & 96.5 & 98.9 & 98.0 & 99.5 \\
 & gR2D2-L &  & 99.3 & 96.2 & 98.2 & 97.0 & 
 99.3 &  & 99.4 & 97.6 & 98.9 & 98.1 & 99.5 \\
 \hline
 & R2D2 &  & 1.79 & 1.80 & 1.96 & 2.45 & 1.36 &  & 1.17 & 1.20 & 1.26 & 1.68 & 0.90 \\
 & HS &  & 1.98 & 2.05 & 2.24 & 2.67 & 1.50 &  & 1.29 & 1.43 & 1.51 & 1.94 & 0.98 \\
AL & GIGG &  & 2.67 & 2.46 & 2.71 & 3.19 & 2.03 &  & 1.83 & 1.71 & 1.87 & 2.25 & 1.39 \\
 & GHS &  & 2.08 & 2.05 & 2.22 & 2.53 & 1.38 &  & 1.35 & 1.44 & 1.49 & 1.77 & 0.91 \\
 & gR2D2-D &  & {\bf 1.65} & {\bf 1.28} & {\bf 1.42} & {\bf 1.53} & {\bf 0.64} &  & {\bf 1.11} & {\bf 0.99} & {\bf 0.92} & {\bf 1.06} & {\bf 0.41} \\
 & gR2D2-L &  & 1.89 & 1.81 & 1.96 & 2.35 & 1.26 &  & 1.25 & 1.24 & 1.27 & 1.62 & 0.82 \\
 \hline
\end{tabular}
\medskip
\begin{tabular}{ccccccccccccccc}
\hline
\multicolumn{3}{c}{$(n,p)=(150, 200)$}& \multicolumn{5}{c}{SNR$=0.5$} && \multicolumn{5}{c}{SNR$=0.7$}\\
 & Scenario &  & 1 & 2 & 3 & 4 & 5 &  & 1 & 2 & 3 & 4 & 5 \\
 \hline
 & R2D2 &  & 99.3 & 98.1 & 98.9 & 98.6 & 99.4 &  & 99.5 & 98.4 & 99.0 & 98.8 & 99.5 \\
 & HS &  & 99.7 & 97.8 & 99.1 & 98.3 & 99.6 &  & 99.8 & 98.3 & 99.3 & 98.7 & 99.8 \\
CP & GIGG &  & 99.1 & 98.9 & 99.0 & 99.0 & 99.1 &  & 99.0 & 98.5 & 98.9 & 98.7 & 98.8 \\
 & GHS &  & 99.4 & 98.1 & 99.4 & 99.0 & 99.5 &  & 99.8 & 98.6 & 99.4 & 99.3 & 99.7 \\
 & gR2D2-D &  & 99.6 & 97.1 & 99.3 & 98.6 & 99.8 &  & 99.8 & 97.9 & 99.5 & 99.2 & 99.8 \\
 & gR2D2-L &  & 99.5 & 98.5 & 99.3 & 98.9 & 99.6 &  & 99.7 & 98.7 & 99.4 & 99.1 & 99.7 \\
 \hline
 & R2D2 &  & 2.16 & 2.28 & 2.50 & 2.64 & 1.57 &  & 1.42 & 1.56 & 1.66 & 1.83 & 1.05 \\
 & HS &  & 1.62 & 1.63 & 1.86 & 1.80 & 1.13 &  & 1.04 & 1.28 & 1.39 & 1.45 & 0.77 \\
AL & GIGG &  & 3.17 & 3.21 & 3.46 & 3.60 & 2.49 &  & 2.31 & 2.40 & 2.55 & 2.72 & 1.82 \\
 & GHS &  & 1.86 & 1.81 & 2.03 & 1.95 & 1.14 &  & 1.21 & 1.33 & 1.42 & 1.42 & 0.77 \\
 & gR2D2-D &  & {\bf 0.74} & {\bf 0.65} & {\bf 0.75} & {\bf 0.68} & {\bf 0.33} &  & {\bf 0.50} & {\bf 0.50} & {\bf 0.52} & {\bf 0.50} & {\bf 0.22} \\
 & gR2D2-L &  & 1.95 & 2.01 & 2.22 & 2.28 & 1.36 &  & 1.30 & 1.40 & 1.49 & 1.60 & 0.91 \\
 \hline
\end{tabular}
\medskip
\begin{tabular}{ccccccccccccccc}
\hline
\multicolumn{3}{c}{$(n,p)=(200, 1000)$}& \multicolumn{5}{c}{SNR$=0.5$} && \multicolumn{5}{c}{SNR$=0.7$}\\
 & Scenario &  & 1 & 2 & 3 & 4 & 5 &  & 1 & 2 & 3 & 4 & 5 \\
 \hline
 & R2D2 &  & 99.9 & 99.7 & 99.7 & 99.7 & 99.9 &  & 99.9 & 99.7 & 99.8 & 99.8 & 99.9 \\
 & HS &  & 99.9 & 99.6 & 99.7 & 99.6 & 99.9 &  & 100 & 99.7 & 99.9 & 99.8 & 100 \\
CP & GIGG &  & 99.9 & 100 & 100 & 100 & 99.9 &  & 100 & 100 & 100 & 100 & 99.9 \\
 & GHS &  & 99.8 & 99.7 & 99.7 & 99.8 & 99.8 &  & 99.8 & 99.7 & 99.8 & 99.7 & 99.8 \\
 & gR2D2-D &  & 100 & 99.7 & 99.8 & 99.8 & 100 &  & 100 & 99.8 & 99.9 & 99.9 & 100 \\
 & gR2D2-L &  & 99.6 & 99.4 & 99.5 & 99.5 & 99.6 &  & 99.7 & 99.5 & 99.6 & 99.5 & 99.8 \\
\hline
 & R2D2 &  & 1.80 & 1.90 & 2.04 & 2.35 & 1.29 &  & 1.18 & 1.29 & 1.42 & 1.60 & 0.87 \\
 & HS &  & 1.17 & 1.33 & 1.48 & 1.76 & 0.72 &  & 0.55 & 0.73 & 0.85 & 1.01 & 0.39 \\
AL & GIGG &  & 2.92 & 2.93 & 3.09 & 3.33 & 2.53 &  & 2.46 & 2.50 & 2.60 & 2.73 & 2.27 \\
 & GHS &  & 1.53 & 1.58 & 1.75 & 2.03 & 1.02 &  & 0.97 & 1.05 & 1.16 & 1.30 & 0.68 \\
 & gR2D2-D &  & {\bf 0.32} & {\bf 0.33} & {\bf 0.35} & {\bf 0.37} & {\bf 0.18} &  & {\bf 0.21} & {\bf 0.23} & {\bf 0.23} & {\bf 0.26} & {\bf 0.12} \\
 & gR2D2-L &  & 1.38 & 1.45 & 1.57 & 1.77 & 0.96 &  & 0.91 & 0.97 & 1.09 & 1.17 & 0.67 \\
 \hline
\end{tabular}
}
\caption{Coverage probability (\%) and average interval length of $95\%$ credible intervals of regression coefficients, averaged over all the coefficients, under five scenarios of true regression coefficients and two values of SNR. The narrowest credible intervals are represented in {\bf bold}.
}
\label{tab:sim-cp}
\end{table}

\section{Real-data analysis}\label{sec:real}
To illustrate the proposed methods, we analyze the Abalone dataset, available at the UCI Machine Learning Repository (\url{https://archive.ics.uci.edu/dataset/1/abalone}). The response variable is the number of rings (denoted by $y_i$), which serves as a proxy for the age of abalone, and we consider seven covariates (denoted by $z_{i1},\ldots,z_{i7}$): the shell’s longest length, diameter perpendicular to the length, height with meat in the shell, total weight of the abalone, weight of the meat, weight of the gut after bleeding, and the shell's dry weight. We focus on male abalone, resulting in $n=1528$ samples. 

To capture potential non-linear effects of each covariate, we employ an additive model with $L=10$ B-spline basis functions, created via the R package \verb+fda+ \citep{fda}, and treat the set of 10 basis functions as a group.
Hence, $G=7$ and $p_g=10$, so that we have $p=70$ covariates in total. 
Then, the model is 
$$
y_i=\sum_{g=1}^G\bx_{ig}^\top \bbeta_g + \varepsilon_i, \ \ \ \ \ i=1,\ldots,n,
$$
where $\bbeta_g=(\beta_{g1},\ldots,\beta_{gL})^\top$, $\bx_{ig}=(\Phi_{g1}(z_{ig}),\ldots, \Phi_{gL}(z_{ig}))^\top $ and $\Phi_{gl}(z)$ is the $l$th B-spline function.

\begin{figure}[htbp!]
\centering
\includegraphics[width=\linewidth]{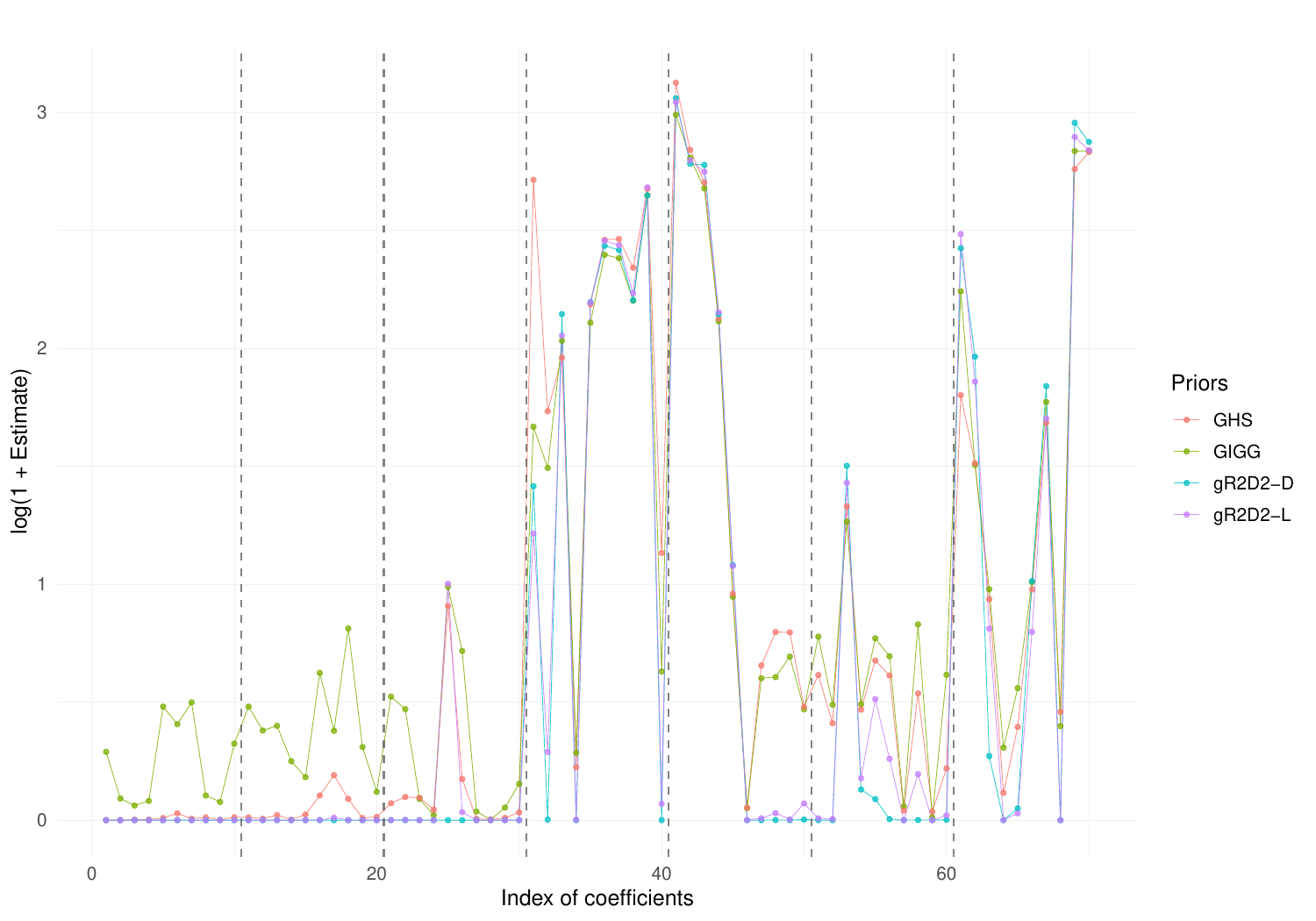}
\caption{Posterior median of regression coefficients obtained by four group shrinkage priors, GHS, GIGG, gR2D2-D and gR2D2-L.} 
\label{fig:app}
\end{figure}

\begin{figure}[htbp!]
\centering
\includegraphics[width=\linewidth]{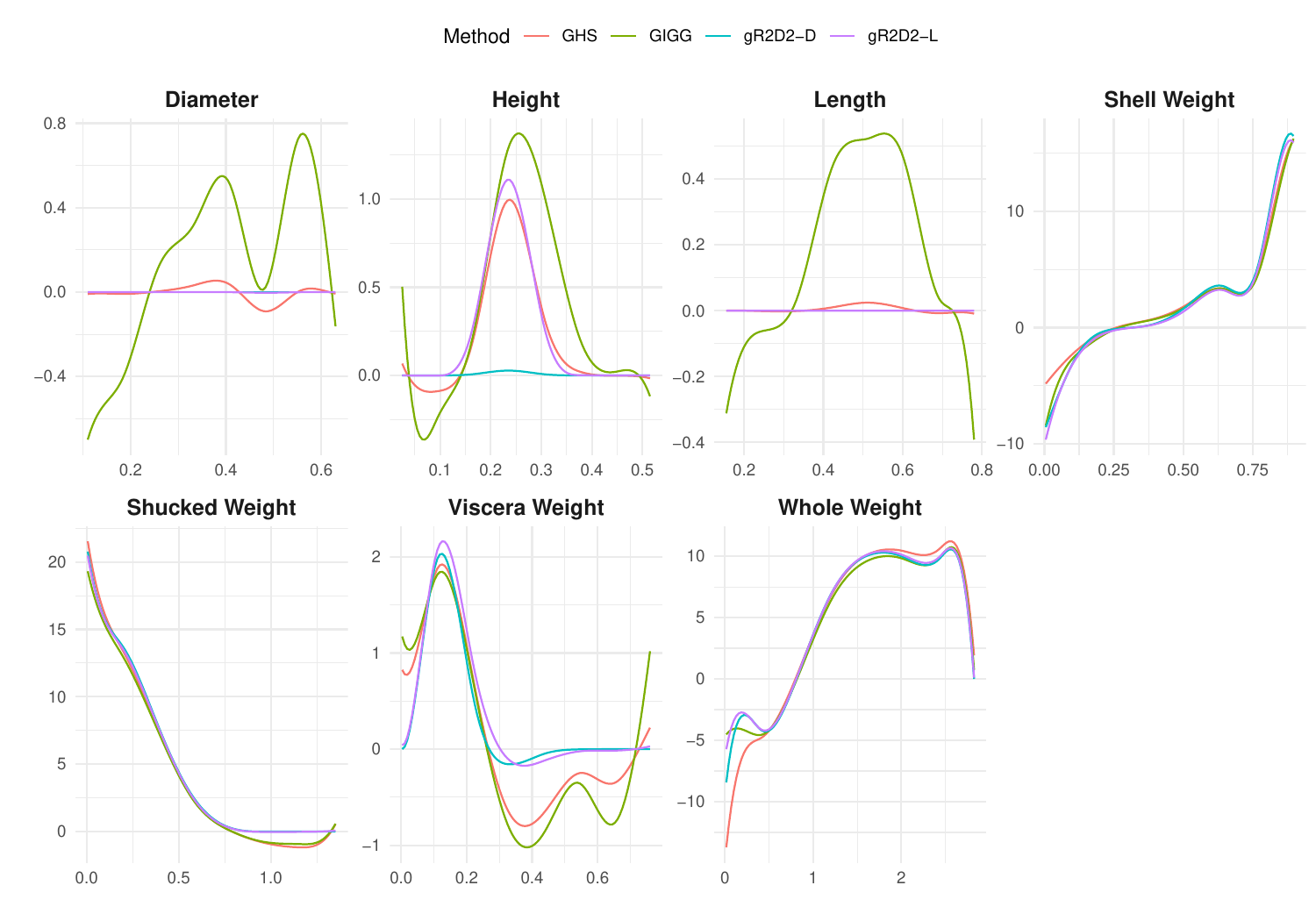}
\caption{Posterior medians of estimated non-linear regression functions of seven covariates obtained by four group shrinkage priors, GHS, GIGG, gR2D2-D and gR2D2-L.} 
\label{fig:app-reg}
\end{figure}

We apply four group shrinkage priors, GHS, GIGG, gR2D2-D and gR2D2-L, as used in the simulation study. 
We generate 3000 posterior samples after discarding the first 1000 samples as burn-in for all the priors. 
The posterior medians of the regression coefficients (coefficients for the basis functions) are presented in Figure~\ref{fig:app}. 
The results show that the proposed gR2D2 priors tend to provide significantly sparser point estimates than those of GHS and GIGG, particularly for the first three functions (i.e., $\bbeta_1, \bbeta_2$ and $\bbeta_3$). This accords with our findings in the simulation study which showed that the gR2D2 prior provides the best estimation of null coefficients. In other words, the proposed prior appears to enforce the most shrinkage towards 0. This is likely because the global variance parameter in the gR2D2 prior has a beta prime distribution which has more prior mass near $W=0$ compared to the half-Cauchy distribution used in the GHS and GIGG priors \citep[e.g.,][]{zhang2022bayesian}. On the other hand, for $\bbeta_4, \bbeta_5$ and $\bbeta_7$, each prior yields much larger estimates.
In Figure~\ref{fig:app-reg}, we plot the posterior median of the non-linear effect for each covariate.
Moreover, we compute $95\%$ credible intervals of non-linear effects, $\sum_{l=1}^L \beta_{gl}\Phi_{gl}(z)$, as a function of $z$, evaluated at 100 equally-spaced points from ${\rm min}_{i=1,\ldots,n}(z_{ig})$ to ${\rm max}_{i=1,\ldots,n}(z_{ig})$. We then compute the point-wise interval lengths and average over the evaluation points.
The average length of each covariate is presented in Table~\ref{tab:app}.
The results indicate that gR2D2 provides much shorter (more efficient) credible intervals than GHS and GIGG, which is consistent with the simulation results in Section~\ref{sec:sim}.

We also evaluate the performance of these priors on out-of-sample predictions. To do so, 500 observations are randomly selected and defined as test data, and the response values are predicted in the test sample by fitting the additive model using the training sample (1028 observations). 
By generating random samples from the posterior predictive distribution for the test data, we calculate mean (point prediction) and $95\%$ prediction intervals of the test data.   
The performance of the point and interval prediction is evaluated through mean squared errors (MSE) and empirical coverage probability (CP), respectively.
This procedure was repeated 100 times and the resulting MSE and CP values are reported in Table~\ref{tab:app}. The MSE and CP are almost identical for all the priors, with the gR2D2 priors yielding slightly lower MSE values.
Combining this with the average length of the non-linear effects results, the proposed gR2D2 prior yields sparser and more efficient results without sacrificing prediction accuracy.
Finally, we notice that all methods have slight under-coverage (around 90\% instead of the prescribed 95\%). While it is difficult to give an exact reason for this, we stress that this is for out-of-sample prediction of $y_i$, not inference on $\bbeta$. The uncertainty of $y_i$ comprises not only $\bbeta$ but also $\varepsilon_i$ (and $\sigma^2$). Thus, with a different prior for $\sigma^2$, i.e., an inverse-gamma prior with an appropriate choice of hyperparameters, the nominal coverage may be achievable.

\begin{table}[htb!]
\centering
\begin{tabular}{ccccccccccccccc}
\hline
&&\multicolumn{7}{c}{Covariate (average length of credible intervals)} & \multicolumn{2}{c}{Out-of-Sample}   \\
 &  & 1 & 2 & 3 & 4 & 5 & 6 & 7 & MSE & CP (\%) \\
 \hline
GHS &  & 3.49 & 4.00 & 15.02 & 10.31 & 5.39 & 5.12 & 7.31 & 5.08 & 90.42 \\
GIGG &  & 1.27 & 1.72 & 8.57 & 11.63 & 7.57 & 4.53 & 9.36 & 5.02 & 90.38 \\
gR2D2-D &  & 0.03 & 0.15 & 2.43 & 7.44 & 3.53 & 3.41 & 6.31 & 4.99 & 90.36 \\
gR2D2-L &  & 0.87 & 1.11 & 7.27 & 8.49 & 3.95 & 4.48 & 6.12 & 5.00 & 90.39 \\
\hline
\end{tabular}
\caption{
Average length of $95\%$ credible intervals of non-linear effects of seven covariates, 1 (Length), 2 (Diameter), 3 (Height), 4 (Whole Weight), 5 (Shucked Weight), 6 (Viscera Weight) and 7 (Shell Weight), and out-of-sample mean squared errors (MSE) and coverage probability (CP) of $95\%$ prediction intervals, obtained from four groups shrinkage priors, GHS, GIGG, gR2D2-D and gR2D2-L.  
}
\label{tab:app}
\end{table}

\section{Conclusion}\label{sec:conc}
In this work, we proposed the gR2D2 prior, a shrinkage prior for grouped regression settings. From first-principles, we derived group-specific $R^2$ values and endowed these values with a Dirichlet distribution. This framework induced a Beta-prime distribution on the group specific variance parameters, as well as dependence between them. We considered both a Dirichlet and logistic normal distribution for the variance allocation parameters and discussed various hyperparameter choices. In particular, we prefer an Empirical Bayes-inspired approach of estimating the group-specific $R^2_g$ to guide the hyperparameter choices. An efficient MCMC algorithm to sample from the posterior demonstrated strong performance on simulated and real-world data.

The simulation study also yielded some insights on when the grouping priors perform well. Not surprisingly, when the signal is equally spread between all the groups, incorporating the group structure into the prior provides minimal benefit, and in some cases actually worsens performance. On the other hand, if one group clearly demonstrates a strong signal, then the proposed approach can account for this in the hyperparameter selection, and will yield good inference properties in point estimation. The importance of modeling the group structure is also seen in posterior uncertainty quantification. We stress that further study is needed to get a clearer picture of when accounting for the grouping structure improves performance.

\section*{Statements and Declarations}
The authors have no competing interests to declare. ChatGPT was used to help generate the R code to produce Figure 3.

\vspace{5mm}
\bibliographystyle{chicago}
\bibliography{refs}

\appendix
\setcounter{equation}{0}
\renewcommand{\theequation}{A\arabic{equation}}

\clearpage

\vspace{0.5cm}
\begin{center}
{\bf {\large Supplemental Materials}}
\end{center}

\section{$R^2$ derivation}
Recall that $\beta_{gj}|\lambda_{gj},\sigma^2$ is such that $\mbox{E}(\beta_{gj})=0$ and $\mbox{Var}(\beta_{gj})=\sigma^2\lambda_{gj}$ with independence among the $\beta_{gj}$'s conditional on the variance parameters, i.e., $\mbox{Var}(\bbeta)=\sigma^2\Lambda$ where $\Lambda$ is a diagonal matrix with $(\lambda_{11},\dots,\lambda_{Gp_G})$ on the diagonal. Furthermore, $\mbox{E}(\bx)={\bf 0}_p$ and $\mbox{Var}(\bx)=\Sigma_x$ where the diagonal of $\Sigma_x$ is all ones. Then we can show:
\begin{align*}
    \mbox{Var}(\bx^\top\bbeta)
    &=\mbox{E}_{\bx}(\mbox{Var}_{\bbeta}(\bbeta|\bx)) + 
    \mbox{Var}_{\bx}(\mbox{E}_{\bbeta}(\bbeta|\bx)) \\
    &=\mbox{E}_{\bx}(\sigma^2\bx^\top\Lambda\bx) + 
    \mbox{Var}_{\bx}(0) \\
    &=\sigma^2\mbox{E}_{\bx}(\mbox{tr}(\bx^\top\Lambda\bx))\\
    &=\sigma^2\mbox{tr}(\Lambda\mbox{E}_{\bx}(\bx\bx^\top))\\
    &=\sigma^2\mbox{tr}(\Lambda\Sigma_x)\\
    &=\sigma^2\sum_{g=1}^G\sum_{j=1}^{p_i}\lambda_{gj}.
\end{align*}
The rest of the derivation of $R^2$ follows immediately.

\section{Proofs of Propositions}

\subsection{Proposition 1}
We start from the $W_g$-based model:
\begin{equation*}
    W_g | \tau ^2 \stackrel{ind}{\sim} \textsf{Gamma}(a_g,\tau^{2}), \qquad \tau ^2 \sim \textsf{InvGamma}(b,1),
\end{equation*}
where we use the shape-scale parameterization of the gamma distribution, i.e., $\mathrm{E}(W_g|\tau^2)=a_g\tau^2$.

The joint density is given by 
\begin{equation*}
\begin{split}
    \pi_{(W_1,\dots ,W_G,\tau^2)}(W_1,\dots , W_G, \tau^2) &= \pi(W_1,\dots , W_G| \tau^2) \pi(\tau^2)  \\
    &=\left[ \prod _{g=1}^G \textsf{Gamma}(W_g|a_g,\tau^{2}) \right] \textsf{InvGamma}(\tau^2|b,1) \\
    &=\left[ \prod _{g=1}^G  \frac{1}{\Gamma (a_g) (\tau^{2} )^{a_g}} W_g^{a_g-1} e^{ -W_g/\tau^2 } \right] \frac{1}{\Gamma(b)} (\tau^2)^{-b-1} e^{ -1/\tau^2} \\
    &= \left[ \prod _{g=1}^G  \frac{1}{\Gamma (a_g)} W_g^{a_g-1} (\tau^2)^{-a_g} e^{ -W_g/\tau^2 } \right] \frac{1}{\Gamma(b)} (\tau^2)^{-b-1} e^{ -1/\tau^2}
\end{split}
\end{equation*}
Then, consider the change of variable, mapping $(W_1,\dots ,W_G,\tau^2 )$ to $(\gamma _1,\dots ,\gamma _G,\tau^2)$ by $W_g = \gamma _g\tau^2$. Note that the Jacobian is $(\tau ^2)^{G}$. Then we have the joint density of $(\gamma_1,\dots ,\gamma_G,\tau^2)$ as,
\begin{equation*}
\begin{split}
    \pi(\gamma _1,\dots , \gamma _G, \tau^2) &= \pi_{(W_1,\dots ,W_G,\tau^2)}(\gamma _1\tau^2,\dots , \gamma _G\tau^2, \tau^2)  \left| \frac{d(W_1,\dots ,W_G,\tau^2)}{d(\gamma _1,\dots ,\gamma_G,\tau^2)} \right| \\
    &= \left[ \prod _{g=1}^G  \frac{1}{\Gamma (a_g)} (\gamma _g \tau^2)^{a_g-1} (\tau^2)^{-a_g} e^{ -(\gamma _g\tau^2)/\tau^2 } \right] \frac{1}{\Gamma(b)} (\tau^2)^{-b-1} e^{ -1/\tau^2} (\tau^2)^{G} \\
    &= \left[ \prod _{g=1}^G  \frac{1}{\Gamma (a_g)} \gamma _g ^{a_g-1} e^{ -\gamma _g } \right] \frac{1}{\Gamma(b)} (\tau^2)^{-b-1} e^{ -1/\tau^2} \\
    &= \left[ \prod _{g=1}^G \textsf{Gamma}(\gamma_g|a_g,1) 
 \right] \textsf{InvGamma}(\tau^2|b,1), 
\end{split}
\end{equation*}
where the new variables are all independent and 
\begin{equation*}
    \gamma _g \stackrel{ind}{\sim} \textsf{Gamma}(a_g,1), \qquad \tau ^2 \sim \textsf{InvGamma}(b,1).
\end{equation*}
Define $W=W_1+\cdots +W_G$.
Then, write the vector of $R^2$'s as
\begin{equation*}
\begin{split}
    (R^2_1,\dots,R_G^2,1-R^2) &= \left( \frac{W_1}{W+1} , \dots, \frac{W_G}{W+1}, \frac{1}{W+1} \right) \\
    &= \left( \frac{\gamma _1}{\sum_{g'=1}^G \gamma_{g'} +\tau^{-2} } , \dots, \frac{\gamma _G}{\sum_{g'=1}^G \gamma_{g'} +\tau^{-2} }, \frac{\tau^{-2}}{\sum_{g'=1}^G \gamma_{g'} +\tau^{-2} } \right) ,
\end{split}
\end{equation*}
which are the independent gamma ratios and follow the Dirichlet distribution.

\subsection{Proposition 2}
The gR2D2 prior is
\begin{multline*}
\beta_{gj}|\phi_{gj}, W_g\sim \mathsf{DE}\left(\phi_{gj} W_g  \right), \quad W_g|\tau^2\sim\mathsf{Gamma}(a_g,\tau^{2}), \quad \\ \tau^2\sim\mathsf{InvGamma}(b,1),
\boldsymbol{\phi}_g\sim\mathsf{Dirichlet}(a_{g1},\dots,a_{gp_g}).
\end{multline*}
If we let $a_{gj}=a_{g\pi}$ for $j=1,\dots,p_g$, then  this is equivalent to
\begin{equation*}
\beta_{gj}|\phi_{gj}, W_g\sim \mathsf{DE}\left(\phi_{gj} W_g  \right), \quad W_g\sim\mathsf{BP}(a_g,b),\ 
\boldsymbol{\phi}_g\sim\mathsf{Dirichlet}(a_{g\pi},\dots,a_{g\pi}).
\end{equation*}
We can see that this is equivalent to the original R2D2 prior so we recover the results from Theorem 1 and Theorem 4 of \cite{zhang2022bayesian} for the tail behavior and origin concentration, respectively.

\subsection{Proposition 3}
We follow the same approach as in \cite{boss2023group}. Again, let $\sigma^2=1$. Then
$$
\bbeta_g|W_g,\psi_{gj},\phi_{gj}\sim\mathsf{N}({\bf 0}_{p_g}, W_g\Phi)
$$
where $\Phi$ is a diagonal matrix with $(\psi_{g1}\phi_{g1},\dots,\psi_{gp_g}\phi_{gp_g})$ on the diagonal. Noting that $W_g|\tau^2\sim\mathsf{Gamma}(a_g,\tau^{-2})$ and integrating out $W_g$, we get
\begin{align*}
    &\pi(\bbeta_g|\psi_{gj},\phi_{gj},\tau^2)\\
    &=\int \pi(\bbeta_g|W_g,\psi_{gj},\phi_{gj},\tau^2) \pi(W_g|\tau^2)\ dW_g\\
    &\propto
    \int_0^\infty W_g^{-p_g/2}\exp\left(-\frac1{2W_g}\bbeta_g^\top \Phi^{-1}\bbeta_g\right) W_g^{a_g-1}\exp(-W_g/\tau^2)\ dW_g\\
    &\propto \int_0^\infty W_g^{a_g-p_g/2-1}\exp\left(- \frac{W_g}{\tau^2}-\frac1{2W_g}\bbeta_g^\top \Phi^{-1}\bbeta_g \right)\ dW_g\\
    &\propto \int_0^\infty V_g^{a_g-p_g/2-1}\exp\left\{- V_g-\frac{\left(\sqrt{2\tau^2\bbeta_g^\top \Phi^{-1}\bbeta_g}\right)^2}{4V_g} \right\}\ dV_g\\
    &\propto 
    \frac{\left(\sqrt{2\tau^2\bbeta_g^\top \Phi^{-1}\bbeta_g}\right)^{a_g-p_g/2}}{\left(\sqrt{2\tau^2\bbeta_g^\top \Phi^{-1}\bbeta_g}\right)^{a_g-p_g/2}}
    \int_0^\infty V_g^{a_g-p_g/2-1}\exp\left\{- V_g-\frac{\left(\sqrt{2\tau^2\bbeta_g^\top \Phi^{-1}\bbeta_g}\right)^2}{4V_g} \right\}\ dV_g\\
    &\propto \left(\sqrt{2\tau^2\bbeta_g^\top \Phi^{-1}\bbeta_g}\right)^{a_g-p_g/2}K_{a_g-p_g/2}\left(\sqrt{2\tau^2\bbeta_g^\top \Phi^{-1}\bbeta_g}\right)
\end{align*}
where we made the change of variables $V_g=W_g/\tau^2$. We recognize this final expression as the kernel of the multivariate Laplace distribution with mean ${\bf 0}_{p_g}$ and covariance $\tau^{-2}\Phi$ \citep{kozubowski2013multivariate}.

\section{Behavior of MCMC algorithms}

We discuss the behavior of the MCMC algorithm for the proposed method. Specifically, we provide acceptance rates under the logistic-normal distribution, traceplots of posterior samples and computation times.

First, the average acceptance rates of $\boldsymbol{\phi}$ for the gR2D2-L prior are reported in Table~\ref{tab:sim-ac}. These results indicate that the acceptance rate is consistently high regardless of dimension and underlying structure of the regression coefficients.

\begin{figure}[htbp!]
\centering
\includegraphics[width=\linewidth]{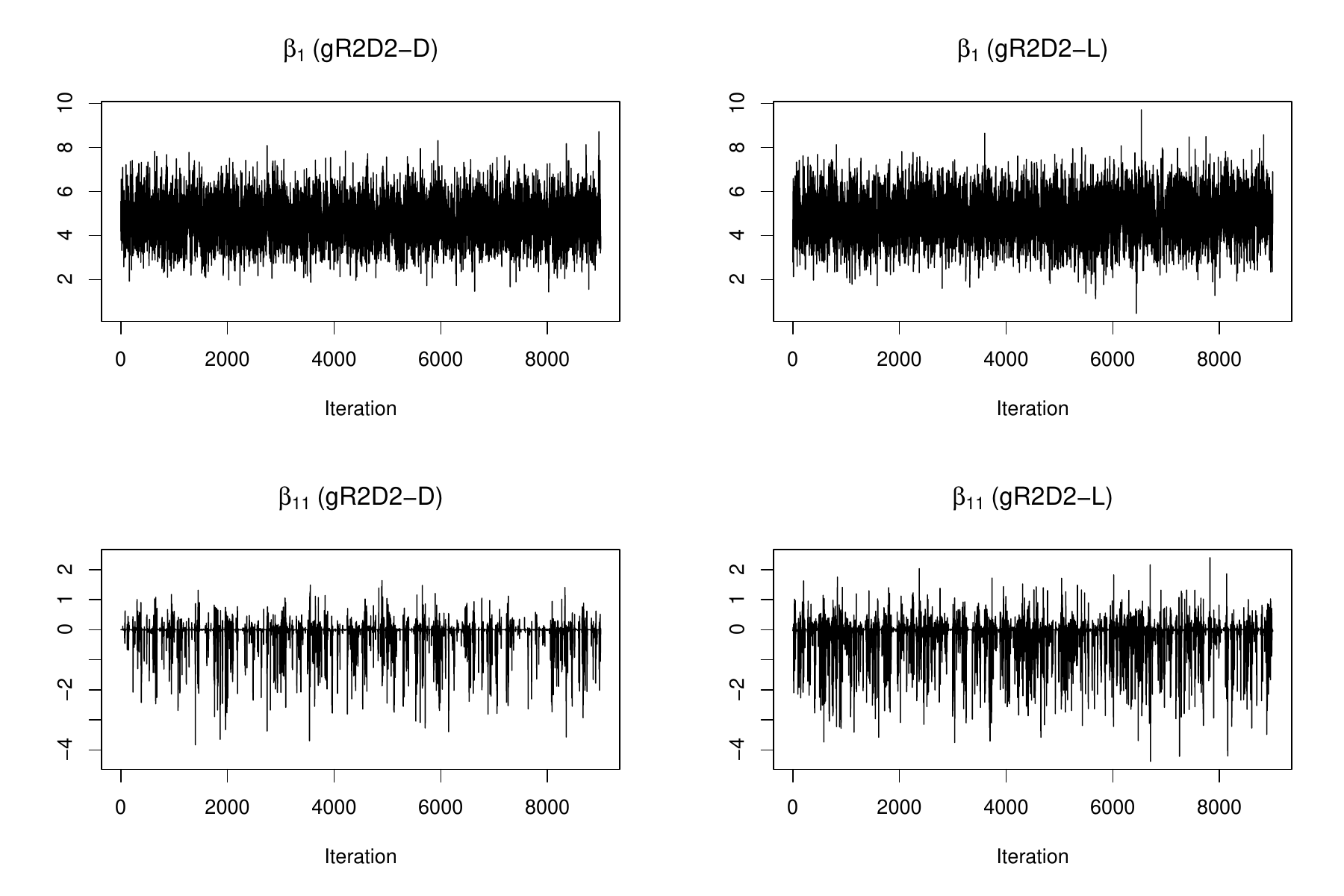}
\caption{Traceplots of $\beta_1$ (non-null) and $\beta_{10}$ (null) of gR2D2-D and gR2D2-L priors, under $(n,p)=(250, 50)$ and Scenario 1. } 
\label{fig:trace1}
\end{figure}

\begin{figure}[htbp!]
\centering
\includegraphics[width=\linewidth]{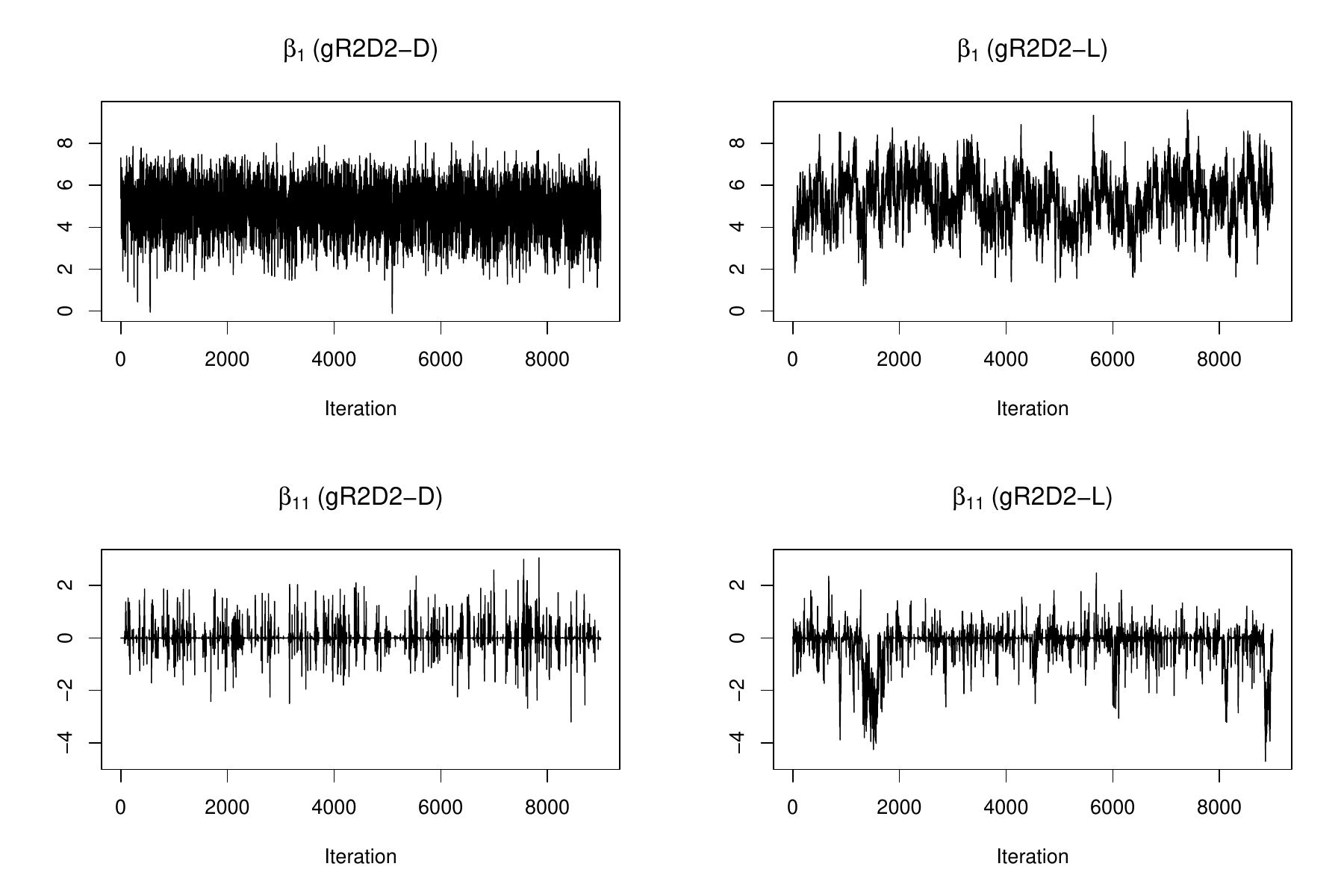}
\caption{Traceplots of $\beta_1$ (non-null) and $\beta_{10}$ (null) of gR2D2-D and gR2D2-L priors, under $(n,p)=(200, 1000)$ and Scenario 1. } 
\label{fig:trace2}
\end{figure}

We next show the traceplots of the posterior samples of the regression coefficients under gR2D2-D and gR2D2-L. 
In Scenario 1, we generate 9000 posterior samples of  $\beta_1$ and $\beta_{10}$ after discarding the first 1000 samples as burn-in. These results are presented in Figures~\ref{fig:trace1} and \ref{fig:trace2} where the true values are $\beta_1=5$ (non-null) and $\beta_{10}=0$ (null). 
For gR2D2-L, the posterior samples seem to be efficiently generated under small dimension ($p=50$), but the serial correlation of the posterior samples seems to be large in higher dimension ($p=1000$).
This may be because the gR2D2-L prior includes a Metropolis-Hasting step in the MCMC algorithm. 
On the other hand, the sampling efficiency of gR2D2-D seems to be less sensitive to the dimension.

Finally, we evaluate the computation time of all priors in Table~\ref{tab:sim-cpt}.
While the computation time of gR2D2 is longer than that of the existing group-level shrinkage priors (GiGG and GHS), the differences are minimal. Moreover, the computation times of gR2D2 are comparable with those of the standard R2D2. As expected, the computation time increases with $p$ for all methods.

\begin{table}[htb!]
\centering
\begin{tabular}{ccccccccccccccc}
\hline
$(n,p)$ && \multicolumn{2}{c}{(250,50)} & \multicolumn{2}{c}{(150,200)} &\multicolumn{2}{c}{(200,1000)} \\
SNR &  & 0.5 & 0.7 & 0.5 & 0.7 & 0.5 & 0.7 \\
\hline
Scenario 1 &  & 83.5 & 82.9 & 84.0 & 83.8 & 82.3 & 82.4 \\
Scenario 2 &  & 84.3 & 83.7 & 84.2 & 84.1 & 82.3 & 82.4 \\
Scenario 3 &  & 84.3 & 83.9 & 84.2 & 84.1 & 82.3 & 82.4 \\
Scenario 4 &  & 84.6 & 84.4 & 84.3 & 84.2 & 82.2 & 82.4 \\
Scenario 5 &  & 84.9 & 84.7 & 84.3 & 84.3 & 82.5 & 82.6 \\
\hline
\end{tabular}
\caption{Acceptance ratio of the Metropolis-Hastings step in gR2D2-L, averaged over 100 Monte Carlo replicates. 
}
\label{tab:sim-ac}
\end{table}

\begin{table}[htb!]
\centering
\begin{tabular}{ccccccccccccccc}
\hline
$(n,p)$ && \multicolumn{2}{c}{(250,50)} & \multicolumn{2}{c}{(150,200)} &\multicolumn{2}{c}{(200,1000)} \\
SNR &  & 0.5 & 0.7 & 0.5 & 0.7 & 0.5 & 0.7 \\
\hline
R2D2 &  & 5.8 & 5.5 & 23.9 & 23.9 & 148.1 & 145.9 \\
HS &  & 3.5 & 3.5 & 15.9 & 15.9 & 108.4 & 108.0 \\
GIGG &  & 3.6 & 3.5 & 16.3 & 16.4 & 111.8 & 110.9 \\
GHS &  & 3.6 & 3.6 & 16.4 & 17.0 & 113.4 & 114.0 \\
gR2D2-D &  & 6.9 & 6.7 & 22.8 & 22.7 & 143.1 & 143.3 \\
gR2D2-L &  & 7.8 & 7.6 & 25.9 & 26.0 & 163.1 & 162.0 \\
\hline
\end{tabular}
\caption{Computation time in Scenario 1.  
}
\label{tab:sim-cpt}
\end{table}

\end{document}